\documentclass [amssymb,amsmath,twocolumn, showpacs]{revtex4-1} 
\usepackage{graphicx,epsfig,amsfonts,amssymb}
\usepackage{bm}
\usepackage{times}
\usepackage{lipsum}
\usepackage{verbatim}

\newcommand{\be}{\begin{equation}}
\newcommand{\ee}{\end{equation}}
\newcommand{\bes}{\begin{subequations}}
\newcommand{\ees}{\end{subequations}}
\newcommand{\bea}{\begin{eqnarray}}
\newcommand{\eea}{\end{eqnarray}}
\newcommand{\ba}{\begin{array}}
\newcommand{\ea}{\end{array}}
\newcommand{\beqn}{\begin{eqnarray*}}
\newcommand{\eeqn}{\end{eqnarray*}}

\newcommand{\la}{\langle}
\newcommand{\ra}{\rangle}

\newcommand{\dg}{\dagger}
\newcommand{\rar}{\rightarrow}

\begin{document}

\title{Constraints on scattering amplitudes in multistate Landau-Zener theory}

\author {Nikolai~A. {Sinitsyn}$^{a}$, Jeffmin Lin$^b$, and  Vladimir Y. Chernyak$^{c}$}
\address{$^a$ Theoretical Division, Los Alamos National Laboratory, Los Alamos, NM 87545,  USA}
\address{$^b$ Department of Mathematics, University of California, Berkeley, CA 94720-3840, USA}
\address{$^c$ Department of Chemistry and Department of Mathematics, Wayne State University, 5101 Cass Ave, Detroit, Michigan 48202, USA}
 
\begin{abstract}
 We derive a set of constraints, which we will call hierarchy constraints (HCs), on  scattering amplitudes of an arbitrary multistate Landau-Zener model (MLZM). The presence of additional symmetries can transform such constraints into nontrivial relations between elements of the transition probability matrix. This observation can be used to derive complete solutions of some MLZMs or, for models that cannot be solved completely, to reduce the number of independent elements of the transition probability matrix.
\end{abstract}
\date{\today}

\maketitle

\section{Introduction}
\label{sec:intro}

Control over quantum matter can be achieved by applying time-dependent fields, whose effects are described by the nonstationary Schr\"odinger equation.
This equation is often hard to explore even numerically because of a lack of conservation laws, strongly oscillatory behavior, and typically large size of a phase space (e.g., for 280 spins-1/2, the size of the state vector exceeds the estimated number of atoms in the observable Universe).
Within MLZM, it is possible to study quantum dynamics in time-dependent fields without approximations. In this model, evolution is described by a Hamiltonian with parameters that change linearly with time \cite{be}:
\begin{equation}
i\frac{d\Psi}{d t} = \hat{H}(t)\Psi, \quad \hat{H}(t) = \hat{A} +\hat{B}t .
\label{mlz}
\end{equation}
Here, $\Psi$ is the state vector in a space of $N$ states; $\hat{A}$ and $\hat{B}$ are constant Hermitian $N\times N$ matrices.  One can always choose the so-called {\it diabatic basis} in which the matrix $\hat{B}$ is diagonal, and if any pair of its elements are degenerate then the corresponding off-diagonal element of the matrix $\hat{A}$ can be set to zero by a time-independent change of the basis, that is
\be
B_{ij}= \delta_{ij}\beta_i, \quad  A_{nm}=0\,\,\, {\rm if} \,\, \beta_{n}=\beta_{m},\,\,n\ne m \in (1,\ldots ,N).
\label{diab3}
\ee
This can be achieved by diagonalizing the matrix $\hat{B}$, followed by (in the case of degeneracy) diagonalizing the projections of the matrix $\hat{A}$ onto the spaces of eigenvectors of $\hat{B}$ that correspond to degenerate eigenvalues of the latter. Constant parameters $\beta_i$ are called the {\it slopes of diabatic levels}. Nonzero off-diagonal elements of the matrix $\hat{A}$ in the diabatic basis are called the {\it coupling constants}. We will denote them by $g_{ij}\equiv A_{ij}$. Diagonal elements of the Hamiltonian
\be
H_{ii}=\beta_i t +\varepsilon_i, \quad \varepsilon_i \equiv A_{ii},
\label{hii}
\ee
are called the {\it diabatic energies}.
Unless specially stated, we will order indexes according to the sizes of corresponding state energies at $t\rar -\infty$, so that for $i>j$ we have $\beta_i>\beta_j$, or if $\beta_i=\beta_j$ then $\varepsilon_i<\varepsilon_j$.

The goal of the multistate Landau-Zener theory is to find the scattering $N\times N$ matrix $\hat{S}$, whose element $S_{nn'}$ is the amplitude of the diabatic state $|n \rangle e^{-i\varphi_n(t)}$ at $t  \rightarrow +\infty$, given that at $t \rightarrow -\infty$ the system was in the diabatic state $|n' \rangle e^{-i\varphi_{n'}(t)}$, where $\varphi_{k}(t)$ is the time-dependent adiabatic phase of the state $|k \rangle$ at $t \rightarrow \pm \infty$, as explained in detail in Appendix~\ref{sec:S-LZ-linear-algebra}. In many applications, only the matrix $\hat{P}$, with elements $P_{nn'}\equiv P_{n' \rar n} \equiv  |S_{nn'}|^2$  called  {\it transition probabilities}, is needed.

Applications of MLZM in mesocopic, atomic, and molecular physics are ubiquitous \cite{atomic,LZ-interferometry,qcontrol,coher,app-spin}. The origin of this model can be traced to the work of Majorana \cite{maj} who generalized any solution for a spin-1/2 in a time-dependent field, using the two-state Landau-Zener-Majorana-St\"uckelberg model \cite{maj,lz,book-LZ} as an example, to arbitrary spin values.

A general analytical solution of MLZM is unknown, but there are many choices of parameters in Eq.~(\ref{mlz}) for which scattering matrices have been found \cite{do,three-state,bow-tie,bow-tie1,reducible,multiparticle,six-LZ,four-LZ,cQED-LZ}. Recently, considerable progress in deriving nontrivial solvable MLZMs has been achieved due to the discovery that if a model satisfies specific integrability conditions its exact analytical solution can be obtained by application of a semiclassical ansatz that corresponds to applying the  solution for two levels  at all pair-wise diabatic level crossings \cite{six-LZ,four-LZ,cQED-LZ}. Currently, integrability conditions and the way to determine transition probabilities  are conjectures, which are not explained but which  are well supported by all known analytically solved models and extensive numerical checks.

In our article we derive a result that explains at least some of the puzzling properties of MLZM. We will argue that, in any model of the form (\ref{mlz}), scattering matrix elements satisfy a set of constraints with hierarchical structure, i.e., the lower level constraints can be used to reduce the number of variables that are connected at higher levels. We will refer to such constraints with the abbreviation ``HC" meaning the {\it hierarchy constraint}.

Going ahead, we formulate the central result. The $M$-th level of the hierarchy ($M<N$) is the expression for the $M\times M$ minor that stays at  the upper left corner of the scattering matrix. For example, the first three HCs read:
\begin{eqnarray}
\label{be1}
\noindent S_{11}&=& e^{- \pi \sum \limits_{k=2}^N |g_{k1}|^2/|\beta_1-\beta_k|}, \\
\nonumber \\
\label{hhhc2}
\noindent  {\rm Det} \left(
\begin{array}{cc}
S_{11} & S_{12} \\
S_{21} & S_{22}
\end{array}
\right) &=&  e^{- \pi \sum \limits_{k=3}^{N} \left( \frac{| g_{k1}|^2}{|\beta_1-\beta_k|}+\frac{ |g_{k2}|^2}{|\beta_2-\beta_k|} \right)},
\end{eqnarray}
\be
\label{hhhc3}
 {\rm Det} \left(
\begin{array}{ccc}
S_{11} & S_{12} & S_{13} \\
S_{21} & S_{22} & S_{23} \\
S_{31} & S_{32} & S_{33}
\end{array}
\right) =e^{- \sum \limits_{k=4}^{N}\left(  \frac{\pi |g_{k1}|^2}{|\beta_1-\beta_k|}+\frac{\pi |g_{k2}|^2}{|\beta_2-\beta_k|}+\frac{\pi|g_{k3}|^2}{|\beta_2-\beta_k|} \right)}.
\ee
There is a second hierarchy that starts with the right lower corner of the scattering matrix. The first two such HCs read
\begin{equation}
\label{be2}
S_{NN}= e^{- \pi \sum \limits_{k=1}^{N-1} |g_{kN}|^2/|\beta_N-\beta_k|},
\end{equation}
\be
\label{hhhc21}
{\rm Det} \left(
\begin{array}{cc}
S_{N-1,N-1}& S_{N-1,N} \\
S_{N,N-1} & S_{NN}
\end{array}
\right) =  e^{-  \sum \limits_{k=1}^{N-2} \left( \frac{\pi| g_{kN}|^2}{\beta_N-\beta_k}+\frac{\pi |g_{k,N-1}|^2}{\beta_{N-1}-\beta_k} \right) }.
\ee
\begin{widetext}
 More generally, HCs are given by
\begin{eqnarray}
\label{h111}
&{\rm Det}& \left(
\begin{array}{cccc}
S_{11}& S_{12}& \cdots & S_{1M} \\
S_{21} & S_{22} & \cdots &S_{2M} \\
\vdots & \cdots  &  \ddots & \vdots \\
S_{M1} & \cdots & \cdots & S_{MM}
\end{array}
\right) =  e^{-\pi \sum \limits_{k=M+1}^{N} \sum \limits_{r=1}^{M} \frac{|g_{kr}|^2}{|\beta_r-\beta_k|} },\quad M =1,\ldots N-1,\\
\label{h222}
\nonumber \\
&{\rm Det}& \left(
\begin{array}{cccc}
S_{N-M+1, N-M+1}& S_{N-M+1,N-M+2}& \cdots & S_{N-M+1,N} \\
S_{N-M+2,N-M+1} & S_{N-M+2,N-M+2} & \cdots &S_{N-M+2,N} \\
\vdots & \cdots  &  \ddots & \vdots \\
S_{N,N-M+1} & \cdots & \cdots & S_{NN}
\end{array}
\right) =  e^{-\pi \sum \limits_{k=1}^{N-M} \sum \limits_{r=N-M+1}^{N} \frac{|g_{kr}|^2}{|\beta_r-\beta_k|} },\quad M =1,\ldots N-1.
\end{eqnarray}
\end{widetext}

The reader familiar with the prior literature on MLZM can recognize that the first level constraints, Eqs.~(\ref{be1})~and~(\ref{be2}), correspond to the known result called the Brundobler-Elser formula that provides the amplitude to remain on a level with an extremal slope. The fact that this formula is only one of a bigger set of exact constraints is our main observation, which has consequences that we will discuss.

The structure of our article is as follows. In Sec.~\ref{derivation}, we derive HCs (\ref{h111})-(\ref{h222}). In Sec.~\ref{prelim}, we review some of the basic information about  MLZM. In Sec.~\ref{chain}, we discuss simple applications that lead to relations between transition probabilities in chain models.  In Sec.~\ref{bands}, we show how HCs imply no-go constraints in MLZM \cite{no-go} and argue that the solution of the Demkov-Osherov model can be derived using only HCs and the unitarity of evolution. In Sec.~\ref{spin32-sec} we prove the validity of the previously conjectured solution of a model with four interacting states. In Sec.~\ref{compare}, we derive the solution of the 4-state generalized bow-tie model using HCs and compare this solution to the result of the application of HCs to a very similar but not fully integrable model. We then discuss our findings in the conclusion, where we also outline open questions.

\section{Derivation of Hierarchy Constraints}
\label{derivation}
The derivation of Eqs.~(\ref{h111})~and~(\ref{h222}) is based on the fact that first level HCs (\ref{be1})~and~(\ref{be2}) are already rigorously proved \cite{mlz-1} (see also \cite{shytov,no-go} for an earlier, more intuitive proof and \cite{joye-LZ,aoki-LZ} for mathematical studies of the Stokes phenomenon in MLZM). Another ingredient is the observation made in \cite{multiparticle} that each model of the type (\ref{mlz}) can be used to generate a bigger model of the form (\ref{mlz}), with the scattering matrix of the bigger model being fully constructed from the scattering matrix of the original model. We formalize this property in Appendix~\ref{sec:LZ-linear-algebra}. For this article, we will only need that with the original matrix Hamiltonian $\hat{H}$ we can associate the secondary quantized Hamiltonian  $\hat{H}'$:
\be
\hat{H}'=\sum \limits_{i,j=1}^N \hat{c}_i^{\dg} H_{ij} \hat{c}_j,
\label{sc}
\ee
where $\hat{c}_i$ and $\hat{c}_i^{\dg}$ are, respectively, annihilation and creation operators of spinless fermions. Note that there are $N$ such operators, i.e., one per diabatic level of the original model. So, one can think about the Hamiltonian (\ref{sc}) as describing hopping of noninteracting fermions among $N$ sites.

The Hamiltonian (\ref{sc}) conserves the number of fermions. For example, if there is only one fermion in the model, the matrix form of the Hamiltonian $\hat{H}'$ coincides with $\hat{H}$. However, if we populate this system with $M>1$ fermions then the matrix form of
$\hat{H}'$ would correspond to evolution of $N!/[M!(N-M)!]$ quantum states. Ref.~\cite{multiparticle} showed that such a Hamiltonian acting in the space of $M$ fermions has the form (\ref{mlz}).  If an arbitrary operator $\hat{X}$ is time independent in the Schr\"odinger picture then in the Heisenberg picture this operator changes with time according to
\be
\frac{d\hat{X}}{dt}= -i[\hat{X}, \hat{H}'].
\label{heis}
\ee
So, in the Heisenberg picture, the Hamiltonian (\ref{sc}) leads to equation
\be
i\frac{d}{dt} \hat{c}_i =\sum_j H_{ij}(t) \hat{c}_j, \quad i,j=1,\ldots, N,
\label{proof1}
\ee
which coincides with the Schr\"odinger equation for state amplitudes in the single particle sector of the model. Since the evolution equation (\ref{proof1}) is linear, we can write the solution for operator evolution from $t\rar -\infty$ to $t\rar +\infty$ in terms of the scattering matrix elements of the single particle sector:
\be
\hat{c}_i (+\infty)=\sum_j S_{ij} \hat{c}_j (-\infty).
\label{proof2}
\ee

For  a sector with $M$ fermions, let indexes $\gamma_1<\gamma_2< \ldots < \gamma_M$ correspond to levels that are initially (at $t\rightarrow -\infty$) populated with fermions, and let $\alpha_1< \alpha_2< \ldots < \alpha_M$ be indexes of the levels populated with fermions at $t\rightarrow +\infty$. Corresponding states are constructed as
\begin{eqnarray}
\label{gamd}
 | \gamma_1, \ldots, \gamma_M \ra &\equiv& \hat{c}^{\dg}_{\gamma_1}(-\infty) \ldots \hat{c}^{\dg}_{\gamma_M}(-\infty) |0 \ra,\\
 \label{ald}
 |  \alpha_1, \ldots, \alpha_M \ra &\equiv& \hat{c}^{\dg}_{\alpha_1}(+\infty) \ldots \hat{c}^{\dg}_{\alpha_M}(+\infty) |0 \ra.
\end{eqnarray}
 Transition amplitudes between such states are given by
\be
S'_{\alpha_1 \ldots \alpha_M, \gamma_1 \ldots \gamma_M } = \la \alpha_1, \ldots, \alpha_M
 | \gamma_1, \ldots, \gamma_M \ra = {\rm Det}(\hat{Q}),
\label{pr1}
\ee
where
\be
\hat{Q} =\left(
\begin{array}{cccc}
S_{\alpha_1\gamma _1} & S_{\alpha_1 \gamma_2} &\ldots & S_{\alpha_1 \gamma_M} \\
S_{\alpha_2 \gamma_1} & S_{\alpha_2\gamma_2} & \ldots & \vdots \\
\vdots &  \vdots & \ddots &\vdots \\
S_{\alpha_M \gamma_1}& \cdots & \cdots& S_{\alpha_M \gamma_M}
\end{array}
\right).
\label{pr4}
\ee

In Appendix~\ref{reduce-ex}, we provide an illustrative example of how one solvable model can produce another model whose solution can be obtained using Eq.~(\ref{pr1}). For this section, however, we do not assume that the single particle sector of the model is exactly solvable. Instead, we note that first level HCs, Eqs.~(\ref{be1}) and (\ref{be2}), can be equally applied to the sector with $M$ fermions. In such a model, the lowest slope diabatic level corresponds to the state in which all fermions occupy the $M$ first lowest slope levels of the original $N$-state system, i.e., $\gamma_k=k$, $k=1,\ldots,M$. This state is coupled directly to and only to states with one of the indexes $\gamma_k$ replaced by some $r$, such that $N\ge r>M$. Corresponding couplings are equal to $\pm g_{rk}$, where $(\pm)$ sign depends on rearrangement of fermion indexes according to ordering of creation operators in the definition of multifermion states in (\ref{gamd})-(\ref{ald}). Corresponding differences of level slopes are equal, in absolute value, to $|\beta_{k}-\beta_{r}|$.

The survival amplitude for such an extremal state is given by Eq.~(\ref{pr1}) with $\gamma_k=\alpha_k=k$, $k=1,\ldots,M$, which coincides with the left hand side of Eq.~(\ref{h111}). On the other hand, according to Eq.~(\ref{be1}), this survival amplitude is given by the exponential of the sum over terms corresponding to directly coupled states to the extremal one, which is written on the right hand side of Eq.~(\ref{h111}). Note that since such an amplitude depends only on the absolute value of the couplings, inevitable different signs, $\pm$, near couplings of the multi-fermion sector do not produce any effect on the survival amplitude.
Combining two expressions for this amplitude, we arrive at the desired Eq.~(\ref{h111}). The second hierarchy (\ref{h222}) is proved analogously by assuming that all $M$ fermions initially occupy $M$ levels with the highest rather than lowest slopes.


The above derivation of  Eqs.~(\ref{h111}) and (\ref{h222}) was achieved by application of the fist level HCs (\ref{be1}) and (\ref{be2}) to the
$M$-fermion Hamiltonian. One may wonder whether more complex cases of Eqs.~(\ref{h111}) and (\ref{h222}) can give rise to even more complex HCs after such constraints are applied to multiparticle sectors. Our test in Appendix~D shows, however, that such HCs are likely not independent.

Despite the abundance of HCs, there is the question of how much they reveal about  transition probabilities. Numerical simulations of MLZM have shown that scattering matrices generally depend on parameters $\varepsilon_i $, defined in (\ref{hii}), and on the phases of coupling constants. In contrast, HCs do not depend explicitly on $\varepsilon_i$ and depend on couplings and level slopes only via combinations $|g_{ij}|^2/|\beta_i-\beta_j|$.
Moreover, unlike Eqs.~(\ref{be1}) and (\ref{be2}), higher order HCs depend on phases of scattering matrix elements nontrivially.  We will show that, nevertheless, HCs can become sufficient to solve a model if this model has additional symmetries.



\section{Preliminary information}
\label{prelim}
\begin{figure}
\scalebox{0.15}[0.15]{\includegraphics{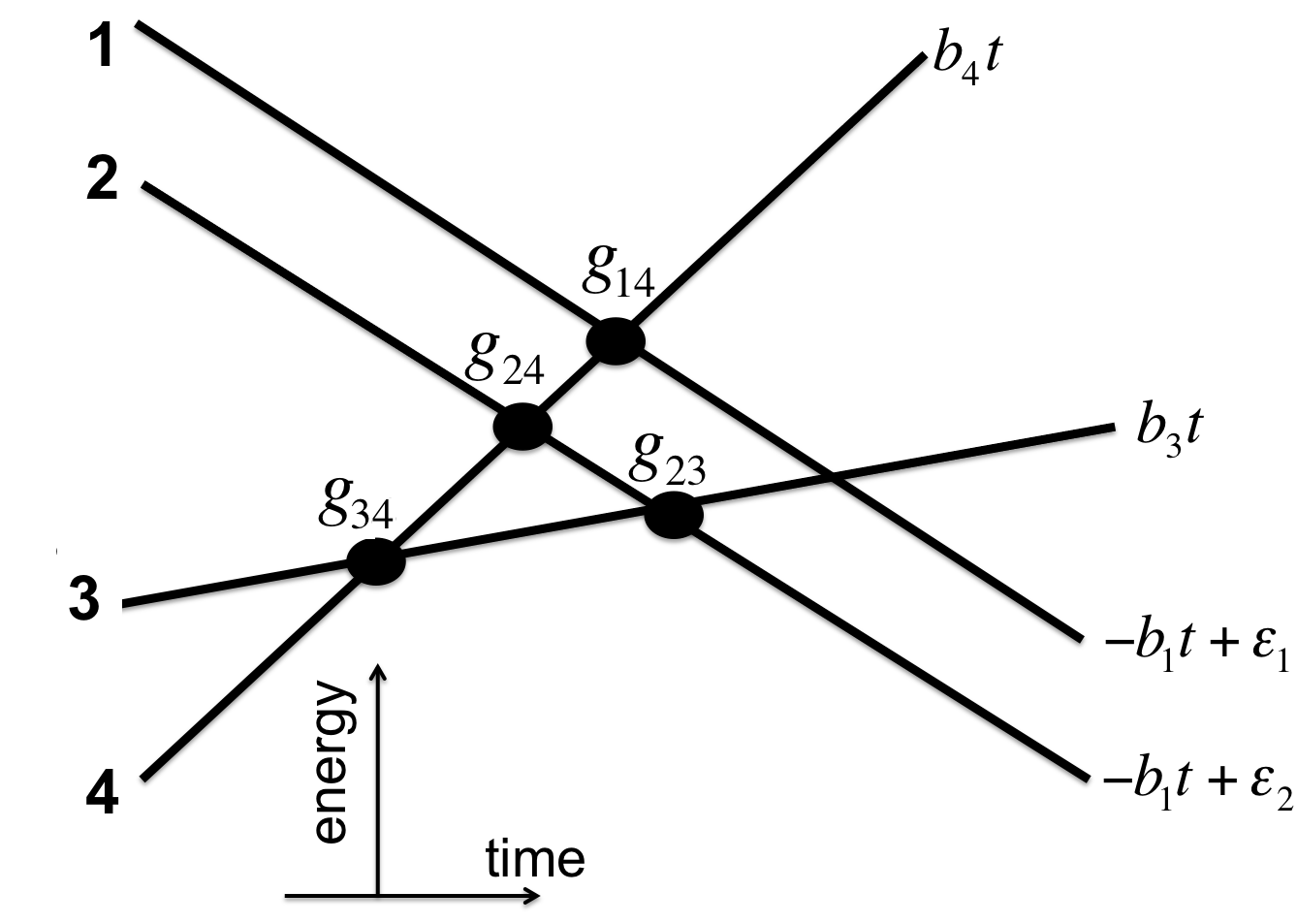}}
\hspace{-2mm}\vspace{-4mm}
\caption{Typical diagram of diabatic levels of a multistate Landau-Zener model. Here, levels 1 and 2 have equal extremal (lowest) slope, and  $\varepsilon_1>\varepsilon_2$.}
\label{gen4-fig}
\end{figure}
Before we proceed with applications, here we review some of the known properties of MLZM. We will then often use this section for references.
\subsection{Diabatic Level Diagram}
It is convenient to illustrate the parameters of any multistate Landau-Zener model on a graph with time-energy axes, as shown in Fig.~\ref{gen4-fig}. Lines of the graph show the time-dependence of diabatic levels (diagonal elements of the Hamiltonian). Small black filled circles mark the intersections of levels with nonzero pair-wise couplings. Integers on the left side of diabatic levels mark level indexes. On the right, levels are marked by analytic expressions for diabatic energies. It is easy to read off the Hamiltonian of the model from such a picture. For example,
we have for Fig.~\ref{gen4-fig}:
\be
\hat{H} =\left(
\begin{array}{cccc}
-b_1 t+\varepsilon_1 & 0&  0& g_{14} \\
0 & -b_1t+\varepsilon_2 & g_{23}& g_{24} \\
0&g_{23}^* & b_3 t &g_{34}  \\
g_{14}^* &g_{24}^* &g_{34}^* & \beta_4 t
\end{array} \right), \quad \varepsilon_1>\varepsilon_2.
\label{bbham1}
\ee
Direct couplings of parallel levels, such as levels 1 and 2 in Fig.~\ref{gen4-fig}, are always considered zero, and if the intersection of two levels is not specially marked then the corresponding coupling is also assumed to be zero, e.g. the coupling between levels 1 and 3 in Fig.~\ref{gen4-fig}.

\subsection{Demkov-Osherov (DO) and bow-tie models}
\label{prelim-anzatz}
\begin{figure}
\scalebox{0.19}[0.19]{\includegraphics{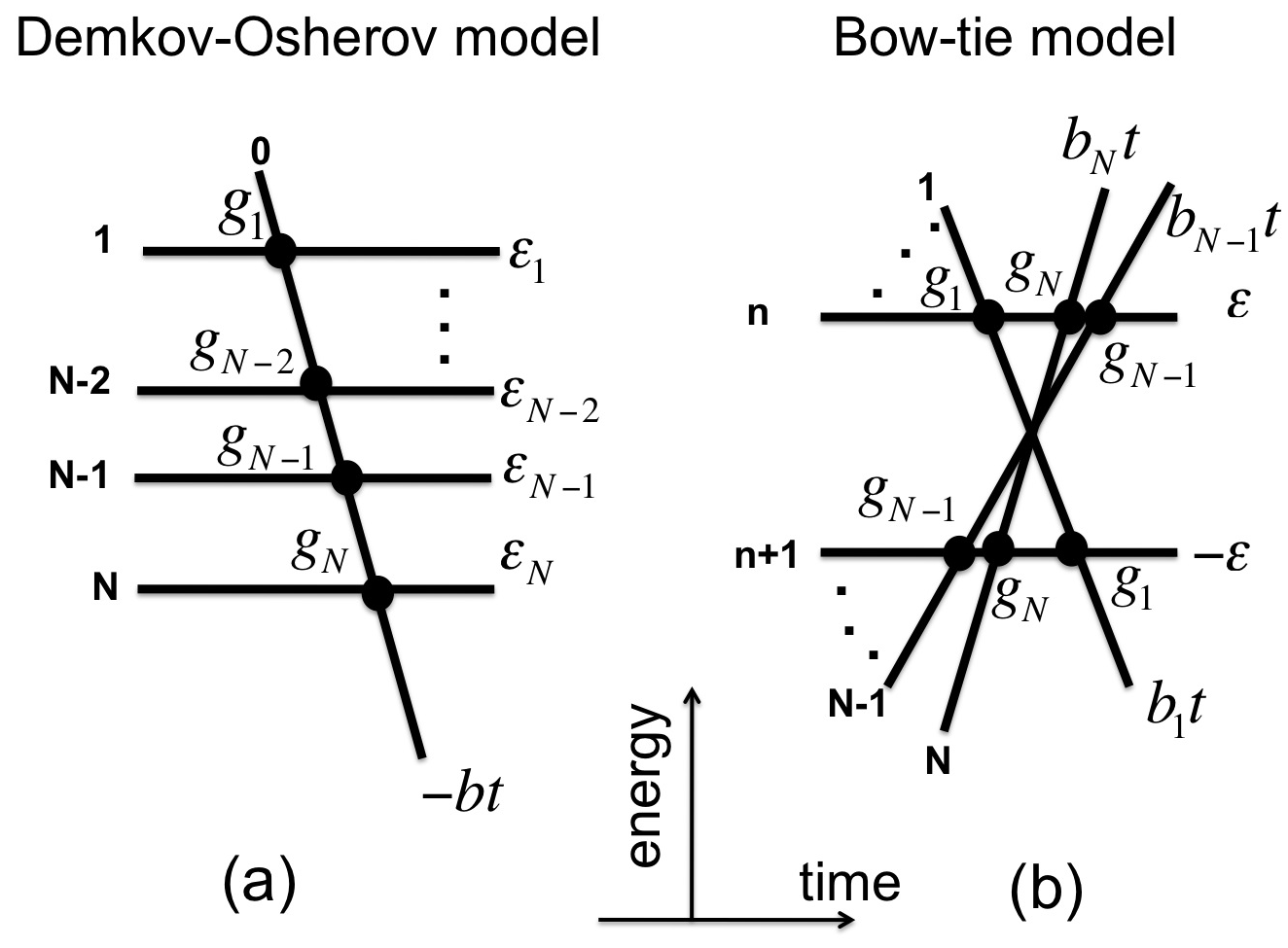}}
\hspace{-2mm}\vspace{-4mm}
\caption{Diabatic level diagrams of  (a) Demkov-Osherov (DO) and (b) bow-tie models. Numbering of levels in DO model  starts with zero.}
\label{do-bt-fig}
\end{figure}

Two models of the type (\ref{mlz}) have been known to be completely solvable for quite some time. One is the Demkov-Osherov (DO) model \cite{do} and another is the bow-tie model \cite{bow-tie,bow-tie1}. Their parameters are illustrated in Fig.~\ref{do-bt-fig}.

The DO model describes the case when a single diabatic level crosses a band of parallel levels (Fig.~\ref{do-bt-fig}(a)). The bow-tie model describes the case when some $N-2$ levels intersect at one point and do not interact with each other directly (Fig.~\ref{do-bt-fig}(b)). Instead, each of them interacts with two parallel levels that are equally distanced from the multilevel crossing point. For any level of the first set, the coupling to each of the parallel levels is the same.

Originally, solutions of these models were found using methods from complex analysis, which could not be applied to any other system of the type (\ref{mlz}). Interestingly, it was observed that, despite the complexity of derivation, transition probabilities in both models are provided by a simple {\it semiclassical ansatz} \cite{bow-tie,six-LZ}. For real valued couplings, the transition probabilities are generated as follows:

1) One should first identify all possible trajectories on a graph in Fig.~\ref{do-bt-fig}(a) or Fig.~\ref{do-bt-fig}(b) that respect causality and connect one initial state and one final state of interest. If there are no such trajectories, the corresponding transition probability is zero.

2) The amplitude of each trajectory is given by a product of simple Landau-Zener passing or turning amplitudes that are encountered along the trajectory: if the diabatic level of a trajectory does not change at the crossing point with coupling $g_i$ and crossing level slopes $\beta_i$ and $\beta_j$, then the trajectory amplitude gains the factor $\sqrt{p_i}$, where $p_i=\exp(-2\pi|g_i|^2/|\beta_{i}-\beta_j|)$. If the trajectory turns at such a crossing point then it gains an amplitude $\pm i\sqrt{1-p_i}$, where sign ($\pm$) is the same as the sign of the coupling constant $g_i$.

3) The final transition probability is obtained by summing the amplitudes of all trajectories that connect the initial state to the final state, and then taking the square of the absolute value of the result.

Within these models, this ansatz can also be used to reconstruct the scattering matrix up to a dynamic phase that is always the same for different interfering trajectories, so that this phase does not influence final transition probabilities, as is discussed in detail in \cite{bow-tie}. We also note that, according to the semiclassical ansatz, solutions of the DO and bow-tie models depend only on combinations of parameters of the form $|g_{i}|^2/|\beta_i-\beta_j|$, just as in the HCs (\ref{h111})-(\ref{h222}).

\subsection{Unitarity conditions}
The scattering matrix is unitary:
\be
\hat{S}\hat{S}^{\dagger} = \hat{S}^{\dagger}\hat{S} = \hat{1}_N.
\label{un1}
\ee
Taking $[\hat{S}^{\dagger} \hat{S} ]_{nn}=[\hat{S}\hat{S}^{\dagger} ]_{nn}=1$ and using the definition $P_{ij}\equiv |S_{ij}|^2$, we find that the unitarity of evolution imposes constraints on transition probabilities:
\be
\sum_{k=1}^N P_{kr} =\sum_{k=1}^N P_{rk}=1,\quad r=1,\ldots, N,
\label{dstoch}
\ee
i.e., the matrix $\hat{P}$ is doubly stochastic. It is easy to verify that one of the equations in (\ref{dstoch}) is dependent on the others. For example, an arbitrary model with four states has in general seven independent constraints of the form (\ref{dstoch}), and hence nine out of sixteen elements of the matrix $\hat{P}$ can be independent. Additional symmetries of a model may reduce the number of independent transition probabilities further, as we will show in the following sections.

\section{Landau-Zener chain models}
\label{chain}
In Landau-Zener chains, all diabatic levels intersect at one point and only pairs of states with adjacent indexes interact with each other directly. One can show that a simple gauge transformation can make all couplings in a chain real \cite{chain}, which we will assume done. So the Sch\"odinger equation for diabatic state amplitudes reads
\be
i\dot{a}_n=b_n t a_n+g_na_{n+1} + g_{n-1} a_{n-1}, \quad n=1,\ldots, N,
\label{chain1}
\ee
where it is also assumed that $g_{0}=g_{N}=0$ to keep dynamics within only $N$ states.

\subsection{Constraints on probabilities in arbitrary chain}
Equation~(\ref{chain1}) does not change if we replace $t \rightarrow -t$ and then change sign of all even indexed amplitudes: $a_{2k}\rar - a_{2k}$. Such discrete symmetries lead to a symmetry of the scattering matrix. Equation that is obtained by  replacement $t\rar -t$ is equivalent to evolution backwards in time, so its scattering matrix is the complex conjugate transpose of the one for the original model \cite{sinitsyn-14pra}. Similarly, a change of sign for some amplitude $a_n$ leads to a change of sign of corresponding scattering matrix elements that contain the index $n$, $S_{mn}$ and $S_{nm}$, with $m\ne n$.  Since  Eq.~(\ref{chain1}) is invariant in the simultaneous application of these two operations, the scattering matrix should be invariant too. This means that the scattering matrix elements of the chain model satisfy additional constraints \cite{sinitsyn-14pra}:
\be
S_{ij}=(-1)^{i+j} S_{ji}^*, \quad i,j=1, \ldots, N.
\label{sym2}
\ee
 Equation~(\ref{sym2}) means that diagonal elements of the scattering matrix are purely real and that the transition probability matrix is symmetric: $P_{ij}=P_{ji}$.
Substituting (\ref{sym2}) into (\ref{hhhc2}) and recalling that couplings for chain models are only between levels with adjacent indexes, we find
\be
S_{11} S_{22} +P_{12}=  e^{-\frac{\pi g_{2}^2} {|b_2-b_3|} }.
\label{hyer2}
\ee
Using (\ref{be1}) we then obtain:
\be
P_{22}\equiv |S_{22}|^2=\left(e^{-\frac{\pi g_{2}^2}{|b_2-b_3|}}-P_{12} \right)^2e^{\frac{2\pi g_{1}^2}{|b_1-b_2|}}.
\label{ch22}
\ee
Although Eq.~(\ref{ch22}) does not fix any of the probabilities separately, it is a nontrivial constraint that relates probabilities $P_{22}$ and $P_{12}$. We checked that Eq.~(\ref{ch22}) holds true for two exactly solvable semi-infinite chain models that were studied in \cite{chain} and for the 4-state chain model that was solved  in \cite{cQED-LZ}.

\subsection{Complete solution of 3-state chain model}
\begin{figure}
\scalebox{0.14}[0.14]{\includegraphics{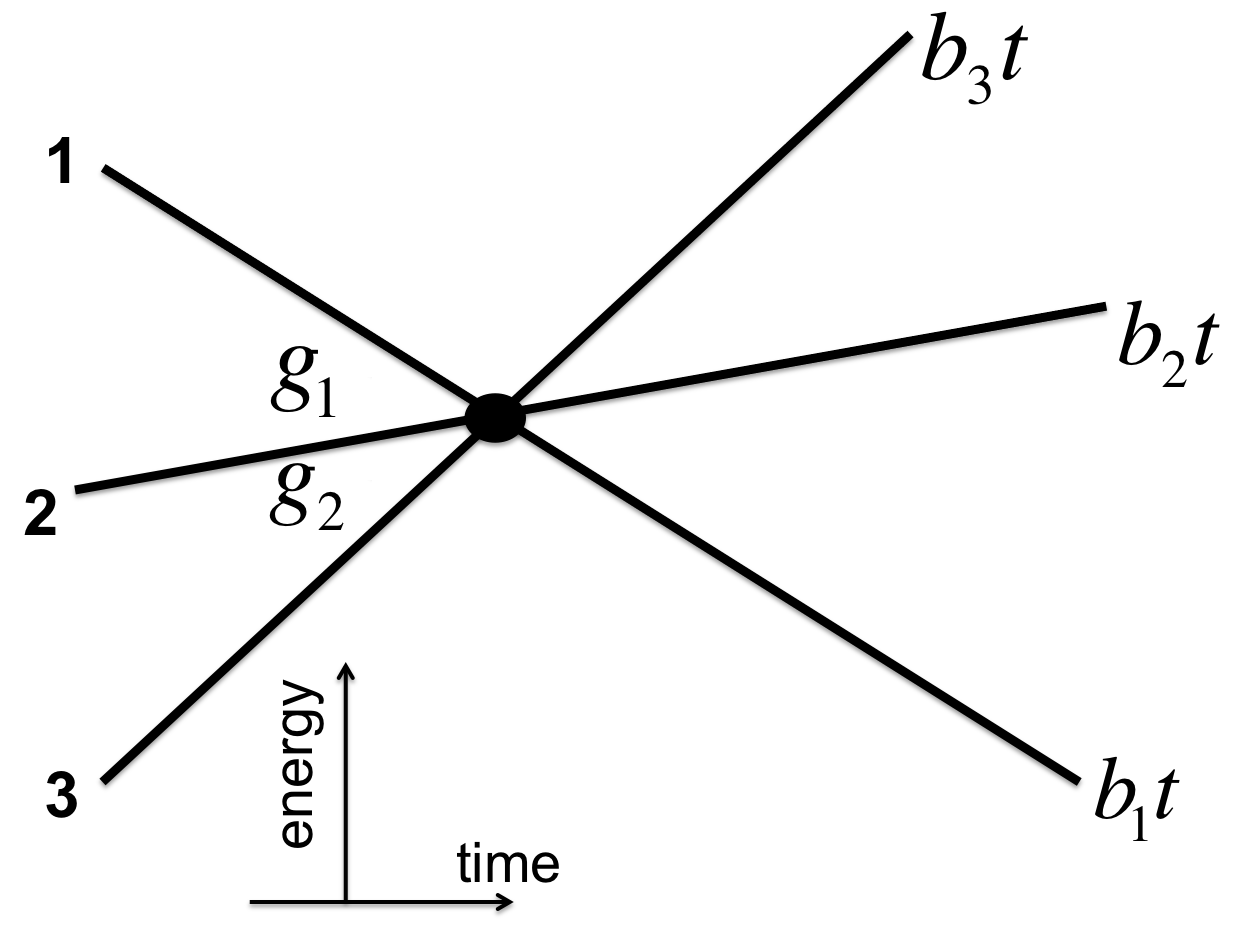}}
\hspace{-2mm}\vspace{-4mm}
\caption{Diabatic level diagram of a three-state chain model. }
\label{three-states}
\end{figure}

One of the earliest known solvable models of the type (\ref{mlz}) is the 3-state chain model. Its Hamiltonian is
\be
\hat{H}=\left(
\begin{array}{ccc}
b_1t & g_{1} &0 \\
g_{1}& b_2t & g_{2} \\
0 & g_{2} & b_3t
\end{array}
\right).
\label{h333}
\ee
It is also illustrated in Fig.~\ref{three-states}. The original solution of this model in \cite{three-state} was very complex. Here we suggest a different approach.

First, the symmetry $P_{ij}=P_{ji}$ that follows from (\ref{sym2}) reduces the number of independent elements of $\hat{P}$ from nine to six. We also recall that this matrix is doubly stochastic. Generally, there are four independent constraints of the type (\ref{dstoch}) for $N=3$, but only three of them are independent for a symmetric matrix. We can use
\begin{eqnarray}
\label{stoch}
\nonumber P_{13}+P_{23} + P_{33}&=&1,\\
P_{11}+P_{12}+ P_{13}&=&1, \\
 \nonumber P_{12}+P_{22}+P_{32}&=&1.
 \end{eqnarray}
 This reduces the number of unknown elements to  three. We then have two HCs (\ref{be1}) and (\ref{be2})
 \be
 P_{11}=e^{-\frac{2\pi g_{1}^2}{|b_1-b_2|}}, \quad  P_{33}=e^{-\frac{2\pi g_{2}^2}{|b_3-b_2|}}.
 \label{be3}
 \ee
Finally, we use constraint (\ref{ch22}), which is specific for chain models.
Altogether we have six equations in (\ref{ch22}), (\ref{stoch}), and (\ref{be3}) for six unknowns: $P_{11},\, P_{22},\, P_{33}, \, P_{12},\, P_{23},$ and $P_{13}$. Solving them, we reproduce the solution in \cite{three-state}, e.g.,
\begin{eqnarray}
P_{22}&=&\left(1-e^{-\frac{\pi g_{1}^2}{|b_1-b_2|}}-e^{-\frac{\pi g_{2}^2}{|b_3-b_2|}} \right)^2,\\
P_{12}&=&\left(1-e^{-\frac{\pi g_{1}^2}{|b_1-b_2|}} \right) \left(e^{-\frac{\pi g_{1}^2}{|b_1-b_2|}}+e^{-\frac{\pi g_{2}^2}{|b_3-b_2|}} \right),
\end{eqnarray}
etc.. Such an algebraic solution is considerably simpler than the solution based on complex analysis of this model in \cite{three-state}.

\section{Band models}
\label{bands}
Here, we will explore the application of HCs to the subclass of models of the type (\ref{mlz}), in which some of the levels have the same slope. We have already shown an example of such a model in Fig.~\ref{gen4-fig}, in which levels 1 and 2 have the same (lowest) slope.

\subsection{No-go rule and extension of Brundobler-Elser formula}
Brundobler and Elser, who noticed Eqs.~(\ref{be1}) and (\ref{be2}) in  numerical simulations \cite{be},  conjectured these formulas only for absolute values, $|S_{11}|$ and $|S_{NN}|$. The first proofs of these formulas in \cite{shytov,no-go} showed, however, that $S_{11}$ and $S_{NN}$ are purely real, so one can drop modulus brackets. Ref.~\cite{no-go} went further and pointed that there are non-rigorous arguments showing that there are two types of extensions of the Brundobler-Elser formula in the case when instead of one level with an extremal slope there is a band of several levels having the same extremal slope but different parameters $\varepsilon_i$ that we defined in (\ref{hii}).

Assume that the band of parallel levels has the lowest slope. First, Ref.~\cite{no-go} suggested that the constraint (\ref{be1}) is applicable to each of the parallel levels:
\be
S_{rr}=e^{-\pi\sum_{k= n+1}^N |g_{kr}|^2/|\beta_k - \beta_n|}, \quad r=1,\ldots, n,
\label{be4}
\ee
where $n$ is the number of levels in the band with extremal slope. Here we recall that parallel levels are assumed not to be directly coupled to each other.

The second suggestion in Ref.~\cite{no-go} was the following no-go rule: for the band with the lowest slope, transitions from level $k$ to level $r$ of the same band (i.e. $1\le k,r \le n$) have zero amplitude if $\varepsilon_k> \varepsilon_r$, i.e.
\be
S_{rk}=0, \quad \varepsilon_k> \varepsilon_r.
\label{no-no}
\ee
There is analogous rule in the case when band levels have the highest slope. Then for levels with $\varepsilon_k> \varepsilon_r$ we have $S_{kr}=0$.

The method suggested in \cite{shytov, no-go} could not be extended rigorously beyond Eq.~(\ref{be1}) for the element $S_{11}$. Eventually, an alternative approach was developed in \cite{mlz-1} that proved both the no-go rule and Eq.~(\ref{be4}). This approach is quite complex. It is based on tedious analysis of perturbation series in powers of coupling constants. Thus, finding a simpler proof of the no-go rule and Eq.~(\ref{be4}) is still desirable. Here we will argue that the no-go rule and Eq.~(\ref{be4}) are consequences of HCs (\ref{h111}).

\begin{figure}
\scalebox{0.15}[0.15]{\includegraphics{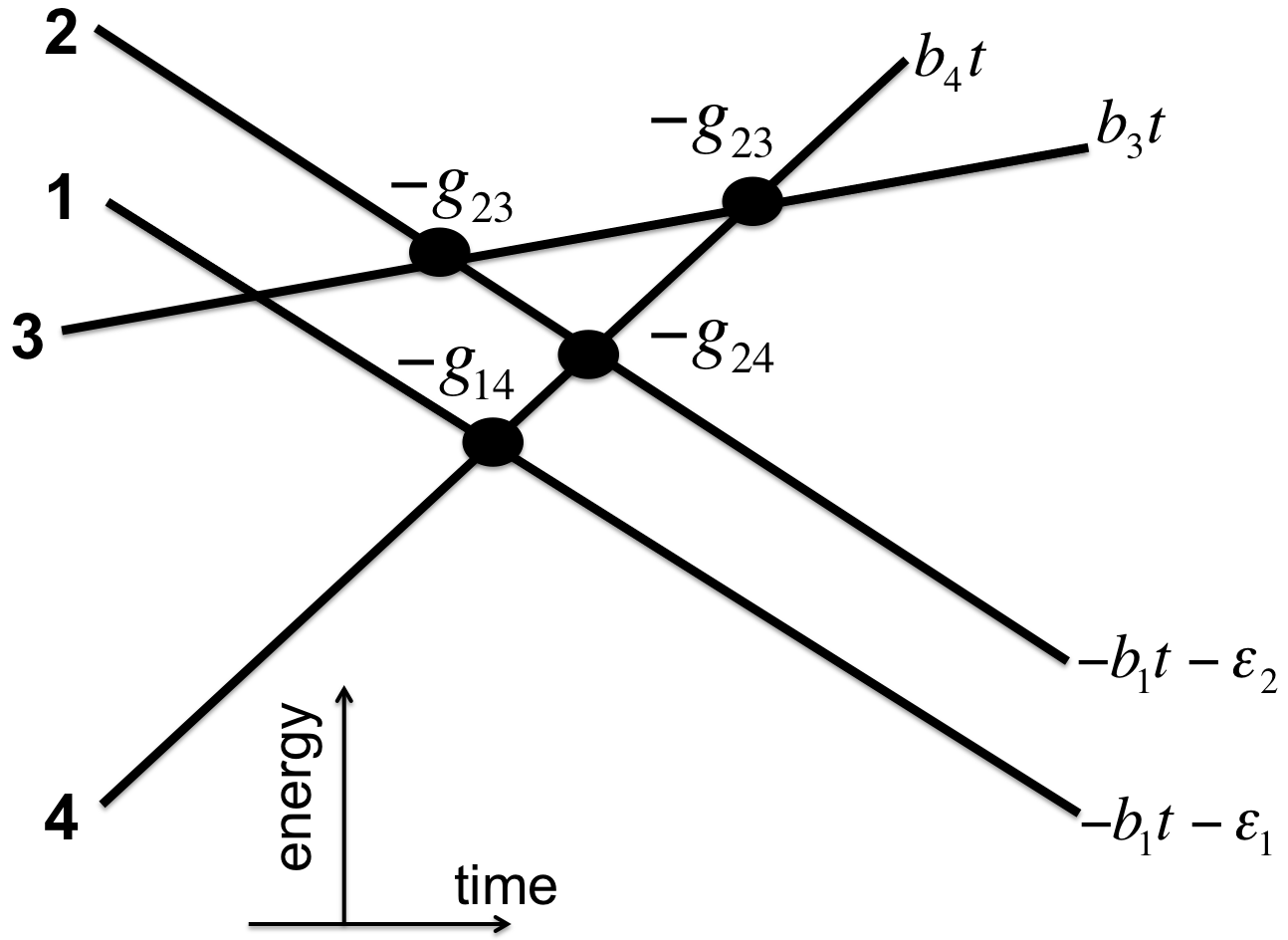}}
\hspace{-2mm}\vspace{-4mm}
\caption{Diabatic levels of the Hamiltonian (\ref{bbham2}) for time-reversed evolution of the model in Fig.~\ref{gen4-fig}. }
\label{four-tau-fig}
\end{figure}

We will illustrate our arguments using the example of the 4-state Hamiltonian $\hat{H}$ in Eq.~(\ref{bbham1}) and Fig.~\ref{gen4-fig}. This Hamiltonian has two parallel levels with the lowest slope.
By our assumptions, Eq.~(\ref{be1})  can be applied to level-1:
\be
S_{11}=e^{-\pi |g_{14}|^2/(b_4-b_1)}.
\label{be6}
\ee
Let us substitute  $t\rar -t$  in the Schr\"odinger equation (\ref{mlz}) with this Hamiltonian. The resulting equation can still be written in the form (\ref{mlz}) but with a new Hamiltonian:
\be
\hat{H}^{\tau} =\left(
\begin{array}{cccc}
-b_1 t-\varepsilon_1 & 0&  0& -g_{14} \\
0 & -b_1t-\varepsilon_2 & -g_{23}& -g_{24} \\
0&-g_{23}^* & b_3 t &-g_{34}  \\
-g_{14}^* &-g_{24}^* &-g_{34}^* & \beta_4 t
\end{array} \right), \quad \varepsilon_1>\varepsilon_2,
\label{bbham2}
\ee
for which the level diagram is shown in Fig.~\ref{four-tau-fig}. Note that the time-reversal operation is equivalent to change of sign of all couplings and parameters $\varepsilon_i$.
Since the Hamiltonian (\ref{bbham2}) describes time-reversed evolution with the original Hamiltonian (\ref{bbham1}), their scattering matrices are related by the conjugate transpose: $\hat{S}^{\tau}=\hat{S}^{\dg}$. This means, in particular, that
\be
 \quad S_{22}^{\tau}=S_{22}^*.
\label{cc1}
\ee

The next observation is that although  matrices (\ref{bbham1}) and (\ref{bbham2}) are related, their corresponding level diagrams in Fig.~\ref{gen4-fig} and Fig.~\ref{four-tau-fig} have different geometries: In Fig.~\ref{four-tau-fig}, level-2 has higher diabatic energy than level-1. Therefore, according to our assumptions, Eq.~(\ref{be1}) can  be applied to level 2 in Fig.~\ref{four-tau-fig}:\
\be
S_{22}^{\tau} = e^{-\pi \left(|g_{23}|^2/(b_3-b_2) + |g_{24}|^2/(b_4-b_2) \right) }.
\label{cc2}
\ee
Substituting this into (\ref{cc1}) we find
\be
S_{22} = e^{-\pi \left(|g_{23}|^2/(b_3-b_2) + |g_{24}|^2/(b_4-b_2) \right) },
\label{cc3}
\ee
which proves that Eq.~(\ref{be4}) works for both parallel levels in this model.

Let us turn to the no-go rule. For  model (\ref{bbham1}) it says that
\be
S_{21}=0.
\label{cc4}
\ee
Let us apply the 2nd level HC to this model:
\be
S_{11}S_{22}-S_{21}S_{12} = e^{-\pi \left( \frac{|g_{14}|^2}{b_4-b_1}+\frac{|g_{23}|^2}{b_3-b_2} + \frac{|g_{24}|^2}{b_4-b_2} \right) }.
\label{cc5}
\ee
Substituting (\ref{cc3}) into (\ref{cc5}) we find
\be
S_{21}S_{12}=0.
\label{cc6}
\ee
This equation can be satisfied either when $S_{21}=0$ or when $S_{12}=0$. Both perturbative and semiclassical calculations would show that $S_{12}\ne 0$ so the only possibility for corresponding parameter ranges is (\ref{cc4}). Since $S_{21}$ is expected to be an analytical function of some combination of parameters, its zero value in a finite interval of parameters means the general validity of (\ref{cc4}).

The above proof of (\ref{be4}) and (\ref{cc4}) is elementary and extendable to
any model (\ref{mlz}) with only two levels in the extremal band. Unfortunately, similar analysis for higher number $n$ has to include an additional nontrivial step that we cannot pursue here in detail: it is expected that transition probabilities are continuous functions of the level slopes if diabatic energies remain non-degenerate except for a few discrete time moments. This should follow from the fact that the time interval with essential nonadiabatic transitions is finite, as one can conclude from analysis of nonadiabatic transitions at large negative and positive time. Therefore, a small variation of level slopes in the Hamiltonian leads to small variations of the corresponding transition probabilities.
If we accept this intuitively obvious but mathematically nontrivial property of scattering amplitudes as true, we can assume that all but two of the band levels with indexes $r$ and $s$ ($s>r$) have slightly lower slopes than other levels of the band, and then end up with (\ref{be4}) for levels $r$ and $s$ and to $S_{sr}=0$. We can then extend this argument to all other pairs of band indexes and thus complete the proof for arbitrary $n$.

\subsection{Next-to-lowest slope band}

As another application of HCs, consider now that there is only one level with the lowest slope and then there are $n$ parallel levels whose slope is the next to the lowest. All other levels are arbitrary except that their slopes are even higher. We are going to derive relations between some combinations of transition probabilities in such models.

Let us ``populate" such a model with two noninteracting fermions by a process described in Sec.~\ref{derivation}. Then the first $n$ levels in the two-fermion model will have the same (lowest) slope,  and we can apply  the no-go rule to them:
\be
S_{1r,1k}=S_{11}S_{rk}-S_{1k}S_{r1}=0, \quad n+1\ge r>k \ge 2,
\label{ng11}
\ee
where $S_{ij,kl}$ and $S_{mn}$ are defined as in Eqs.~(\ref{pr1})-(\ref{pr4}).
Moving one of the terms in (\ref{ng11}) into the right hand side and equating the absolute value squared in both sides of equation we find
\be
P_{11}P_{rk}=P_{1k}P_{r1}, \quad n+1\ge r>k \ge 2,
\label{ng21}
\ee
where $P_{11}$ is known due to (\ref{be1}).

As in the case of a similar relation (\ref{ch22}) in the chain model, Eq.~(\ref{ng21}) does not provide an explicit value of any transition probability. However, it is a nontrivial constraint on the transition probability matrix that reduces the number of its independent parameters.

\subsection{Demkov-Osherov (DO) solution as consequence of HCs}

Consider the DO model in which one level with index $0$ has a slope $-b<0$. This level crosses a band of $N$ parallel levels with zero slopes, as in Fig~\ref{do-bt-fig}(a). The band levels are enumerated so that for $i<j$, we have $\varepsilon_i>\varepsilon_j$. No other diabatic states are present. This model is special because all its levels are extremal, and we can apply (\ref{be4}) to all of them. Let us denote
\be
p_k \equiv e^{-2\pi |g_{k}|^2/b}, \quad q_k=1-p_k, \quad k=1,\ldots, N.
\label{not1}
\ee
Then (\ref{be4}) implies
\be
P_{00}=\prod_{k=1}^N p_k, \quad P_{kk}=p_k.
\label{do-sol1}
\ee
In addition, the no-go constraints for levels of the band, which have the highest slope here, read:
\be
P_{rk}=0, \quad 1\le r < k \le N.
\label{do-sol2}
\ee

Using the semiclassical ansatz in Sec.~\ref{prelim-anzatz}, we can reconstruct other transition probabilities:
\be
P_{m0}=q_m \prod \limits_{k=1}^{m-1} p_k, \quad P_{0m}=q_m \prod \limits_{k=m+1}^N p_k,
\label{do-sol3}
\ee
\be
P_{mn}=q_m q_n \prod \limits_{k=n+1}^{m-1} p_k, \quad m>n>0.
\label{do-sol4}
\ee

Semiclassical ansatz, however, is  a conjecture itself. So, we would rather not use it for a mathematical proof yet. Instead, we are going to  show now  that Eqs.~(\ref{do-sol3})-(\ref{do-sol4}) are also consequences of the HCs. It is easy to check that the solution (\ref{do-sol1})-(\ref{do-sol4}) satisfies the constraints (\ref{ng21}). We count the number of constraints given by Eqs.~(\ref{ng21})-(\ref{do-sol2}) in addition to the constraints (\ref{dstoch}) on the transition probabilities: For $N$ levels in the band, there are $N(N-1)/2$  constraints (\ref{ng21}).  We have also $N(N-1)/2$ no-go rules (\ref{do-sol2}), and $N+1$ rules of the type (\ref{do-sol1}). The rules (\ref{dstoch}) give $2(N+1)-1$ independent constraints for $N+1$ levels. Thus the total number of such constraints is $N^2+2N+2$, which is one larger than the number $(N+1)^2$ of matrix elements of the transition probability matrix. Such a counting, however, does not prove that we have a sufficient number of {\it independent} constraints. To show that we have enough of them, we will construct the solution (\ref{do-sol3})-(\ref{do-sol4}) explicitly, starting from Eqs.~(\ref{ng21})-(\ref{do-sol2}).

Note that (\ref{do-sol4}) is a consequence of (\ref{do-sol3}) and (\ref{ng21}): (\ref{ng21}) implies that for $m > n > 0$, $P_{mn} = P_{0n}P_{m0}/P_{00}$. Performing the substitutions given by (\ref{do-sol3}) directly implies (\ref{do-sol4}). Hence, we are left with showing that (\ref{do-sol3}) is a result of Eqs.~(\ref{ng21})-(\ref{do-sol2}).

The constraints given by Eqs.~(\ref{ng21})-(\ref{do-sol2}) impose the following structure on the $\hat{P}$-matrix of the DO model:
\be
\hat{P}=\left(
\begin{array}{cccccc}
P_{00} & P_{01} &P_{02} & P_{03} & \cdots &  P_{0N}\\
\\
P_{10} & p_1 & 0 & 0 & \cdots & 0 \\
\\
P_{20} & \frac{P_{20}P_{01}}{P_{00}} & p_2 & 0 & \ldots & 0 \\
\\
P_{30} & \frac{P_{30}P_{01}}{P_{00}} & \frac{P_{30}P_{02}}{P_{00}}  & p_3 & \cdots &0 \\
\vdots & \vdots & \vdots & \vdots & \ddots & \vdots \\
P_{N0} & \frac{P_{N0}P_{01}}{P_{00}} & \frac{P_{N0}P_{02}}{P_{00}} & \frac{P_{N0}P_{03}}{P_{00}} &\cdots & p_N
\end{array}
\right).
\label{pr-do1}
\ee
As numbering of levels in the DO model starts with zero, we accept the convention for this model that the numbering of columns and rows of the matrix (\ref{pr-do1}) also starts with zero.

The parameters $p_k$ and $P_{00}$ here are considered known due to Eqs.~(\ref{not1})-(\ref{do-sol1}). Therefore, matrix (\ref{pr-do1}) is  parametrized by $2N$ unknown elements $P_{0k}$ and $P_{k0}$, for $k=1, \ldots, N$. The only constraints that are left express the property that the sum of elements of any row or any column of $\hat{P}$ are equal to one. Applying this property to row 1 (containing $P_{10}$) and column $N$ (containing $P_{0N}$) we find
\be
P_{0N}=q_N, \quad P_{10}=q_1.
\label{re-do2}
\ee
The expressions in (\ref{re-do2}) coincide with the prediction (\ref{do-sol3}) for the same probabilities. Next, let us equate the sum of elements of column 1 or row $N$ to one. Using that $\sum_{k=2}^N  P_{k0} =1-P_{00}-P_{10}$ and $\sum_{k=1}^{N-1} P_{0k} = 1-P_{00} - P_{0N}$ and then rearranging terms, we find
\be
P_{01}= \frac{P_{00}q_1}{1-P_{10}}, \quad P_{N0}= \frac{P_{00}q_N}{1-P_{0N}}.
\label{re-do3}
\ee
Since the elements $P_{0N}$ and $P_{10}$ are known from (\ref{re-do2}), we conclude that we can also derive $P_{01}$ and $P_{N0}$ explicitly.

We continue inductively: look at column $m$ and row $(N - m + 1)$. The constraint for the sum of the elements in this column and row give the expressions
\be
P_{0m}= \frac{P_{00}q_m}{1-\sum \limits_{k=1}^m P_{k0}}, \,\,\,P_{N-m+1,0}= \frac{P_{00}q_{N-m+1}}{1-\sum \limits_{k=1}^m P_{0,N-k+1}}.
\label{re-do5}
\ee
In particular, setting $m=N-1$ we  find from (\ref{re-do5})
\be
P_{0,N-1}= \frac{P_{00}q_{N-1}}{P_{00}+P_{N0}}, \,\,\,P_{20}= \frac{P_{00}q_{2}} {P_{00}+ P_{01}}.
\label{re-do6}
\ee
Now, since $P_{N0}$ and $P_{01}$ are considered known from (\ref{re-do3}), elements $P_{20}$ and $P_{0,N-1}$ can be considered also known due to Eq.~(\ref{re-do6}). In turn, we can substitute their values in (\ref{re-do5}) with $m = 2$ and obtain $P_{02}$ and $P_{N-1,0}$, and so on.

In general, the inductive procedure is as follows: given the values for $P_{01}$ through $P_{0,k-1}$, $P_{0,N-k+2}$ through $P_{0N}$, $P_{10}$ through $P_{k-1,0}$, and $P_{N - k + 2,0}$ through $P_{N0}$, we can use (\ref{re-do5}) with $m = N - k + 1$ to find $P_{0,N - k + 1}$ and $P_{k0}$. Then using (\ref{re-do5}) with $m = k$ gives $P_{0k}$ and $P_{N - k + 1,0}$, continuing the induction.

Consequently, Eqs.~(\ref{ng21})-(\ref{do-sol2}) can be solved recursively reproducing Eqs.~(\ref{do-sol3})-(\ref{do-sol4}). So, indeed, the simplicity of transition probabilities in the DO model is purely the consequence of HCs.


\section{Complete solution of spin-3/2 model}
\label{spin32-sec}
In \cite{six-LZ,four-LZ,cQED-LZ}, several models of the type (\ref{mlz}) were identified that satisfied specific integrability conditions. These models were solved  with the semiclassical ansatz that we described in Sec.~\ref{prelim-anzatz}. Despite definite agreement with the results of numerical simulations, none of these solutions has been rigorously proved analytically yet. Here we consider the simplest of such models that we take from Ref.~\cite{four-LZ} in order to show that HCs, indeed, can be responsible for integrability in this broad class of solvable models.

\begin{figure}
\scalebox{0.14}[0.14]{\includegraphics{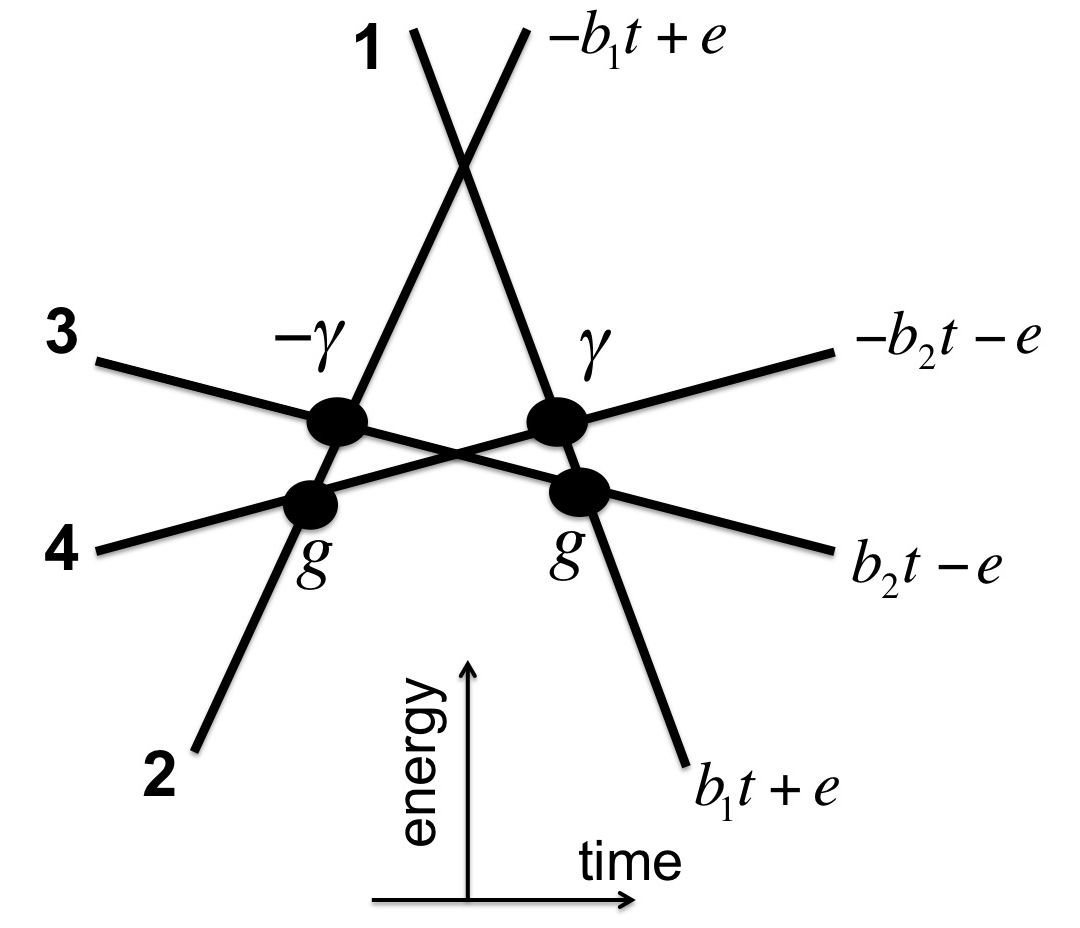}}
\hspace{-2mm}\vspace{-4mm}
\caption{Diabatic levels and couplings of the spin-3/2 model. }
\label{spin32-fig}
\end{figure}
The Hamiltonian of our model is a 4$\times$4 matrix:
\begin{equation}
\hat{H}=\left(
\begin{array}{cccc}
b_1 t +e & 0 & g & \gamma  \\
0 & -b_1 t +e & -\gamma & g  \\
g & -\gamma & b_2 t-e & 0 \\
\gamma&g& 0& -b_2 t -e
\end{array}
\right).
\label{ham44}
\end{equation}
Its structure is explained in Fig.~\ref{spin32-fig}. Note that the numbering of levels is different here from our  convention in order to be consistent with the notation of Ref.~\cite{four-LZ}. The name ``spin-3/2 model" is chosen because this model describes the experimentally relevant situation \cite{app-spin} of a spin-3/2 experiencing quadratic anisotropy and linearly time-dependent magnetic fields, as is explained in \cite{sinitsyn-chemLZ}.

Let us introduce time-dependent amplitudes of the four diabatic states: $a_1(t),a_2(t),a_3(t),a_4(t)$. It was shown in  \cite{four-LZ} that the Schr\"odinger equation (\ref{mlz}) with the Hamiltonian (\ref{ham44}) remains invariant after the simultaneous application of three mutually commuting operations:

(a) time reversal, i.e. the change of $ t \rightarrow -t$, as well as $a_i \rightarrow a_i^*$, $i=1,2,3,4$;

(b) complex conjugation, i.e., the change of the sign near the imaginary unit $i \rightarrow -i$ and replacing  $a_i \rightarrow a_i^*$;

(c) parity operation, i.e. renaming the amplitudes $a_1 \rightarrow -a_2$, $a_2 \rightarrow a_1$, and $a_3 \rightarrow -a_4$, $a_4 \rightarrow a_3$.


As a result, the scattering matrix can be parametrized as shown in \cite{four-LZ}:
\be
\hat{S} = \left(
\begin{array}{cccc}
S_{11}&0&S_{13} &S_{14} \\
0 & S_{11}& S_{23}& S_{24} \\
S_{24}&-S_{14} & S_{44} &0\\
-S_{23} & S_{13} &0 & S_{44}
\end{array}
\right).
\label{scatt3}
\ee
Comparing the off-diagonal elements in (\ref{un1}) with $\hat{S}$ from (\ref{scatt3}) we find relations such as
\be
(\hat{S} \hat{S}^{\dagger})_{13}=S_{11}S^*_{24} +S_{13}S^*_{44}=0.
\label{pr3}
\ee
Based on these relations and using Eq.~(\ref{be1}), Ref.~\cite{four-LZ}
derived relations for transition probabilities:
\begin{eqnarray}
\label{p-diag1}
\nonumber P_{11}&=&P_{22}=P_{33}=P_{44}=p_1p_2,\\
\label{p-diag2}
P_{13}&=&P_{24}=P_{31}=P_{42},\\
\label{p-diag3}
\nonumber P_{14}&=&P_{23}=P_{32}=P_{41},\\
\label{p-diag4}
\nonumber P_{12}&=&P_{21}=P_{34}=P_{43}=0,
\end{eqnarray}
where
\be
p_1\equiv e^{-\frac{2\pi g^2}{|b_1-b_2|}}, \quad p_2\equiv e^{-\frac{2\pi \gamma^2}{|b_1+b_2|}}.
\label{p1p2}
\ee
Equations~(\ref{dstoch}) and (\ref{p-diag2}) fix the transition probabilities up to one unknown parameter. We now show that a 2nd order HC fixes the value of this  parameter.

For the Hamiltonian (\ref{ham44}), we take the extremal level 1 and the next-to-lowest slope level 3. Combining the 2nd order HC (\ref{hhhc2}) with Eq.~(\ref{scatt3}) we find
\be
S_{11}S_{44}-S_{13}S_{24}=p_2.
\label{pr2-3}
\ee
Isolating $S_{24}$ in (\ref{pr2-3}), then substituting the result in (\ref{pr3}), and then using that $P_{11}\equiv |S_{11}|^2$ is given in
(\ref{p-diag1}), we find the missing element of the transition probability matrix:
\be
P_{13}\equiv  |S_{13}|^2 = p_2q_1.
\label{pr2-4}
\ee
Now using the unitarity of the scattering matrix, e.g. $P_{11}+P_{12}+P_{13}+P_{14}=1$, and Eqs.~(\ref{p-diag1}) and (\ref{pr2-4}), we find
\be
P_{14}= q_1.
\label{pr2-5}
\ee
With (\ref{p-diag1})-(\ref{p-diag4}) we can then reconstruct all other elements of the transition probability matrix. This result coincides with the analytical expression suggested in \cite{four-LZ}. Thus the solution in \cite{four-LZ} can now be considered rigorously proven.

Finally, we note that the two-fermion version of this model turns out to belong to the class of  the bow-tie model, as we discuss in  Appendix~\ref{reduction}. This relation can be used to derive not only probabilities but also phases of scattering amplitudes in  model (\ref{ham44}).

\section{Integrable vs nonintegrable Landau-Zener models}
\label{compare}
In this section, we consider two relatively simple models: one is fully solvable and another is not. Both models are intentionally chosen to look very similar, including the presence of a simple discrete symmetry. We will pursue three goals: First, we will keep testing the hypothesis that HCs supplemented by other symmetries are completely responsible for the integrability of known solvable models. Second, we will explore how much information one can extract using HCs about the transition probability matrix of a nonintegrable model. Finally, by comparing very similar integrable and non-integrable models, we identify critical properties of the scattering matrix that lead to full integrability.


\subsection{4-state bow-tie model}
\label{4btm}
\begin{figure}
\scalebox{0.1}[0.1]{\includegraphics{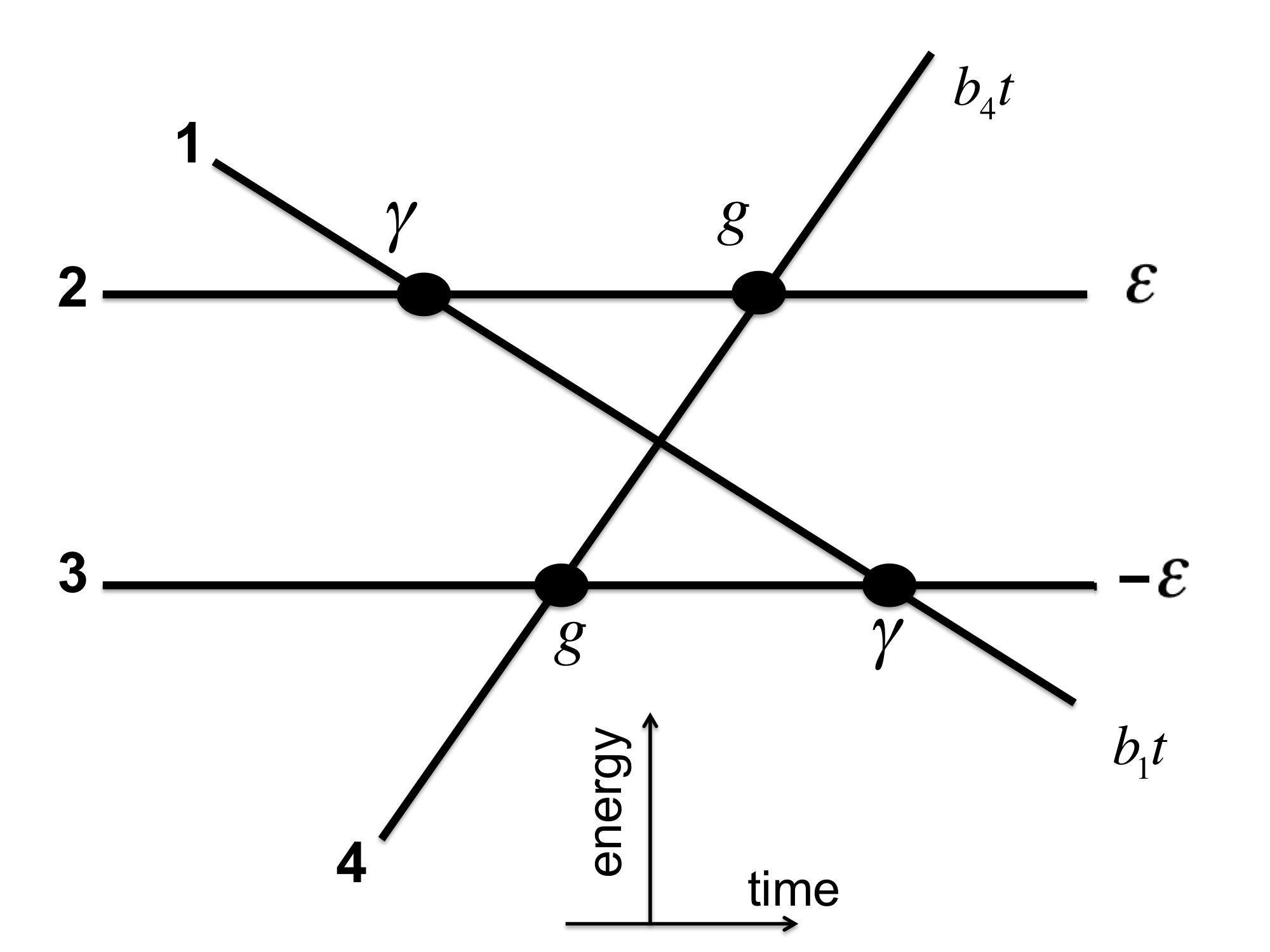}}
\hspace{-2mm}\vspace{-4mm}
\caption{ Diabatic level diagram of the Hamiltonian (\ref{bham1}). }
\label{bow-tie}
\end{figure}

\begin{figure}
\scalebox{0.28}[0.28]{\includegraphics{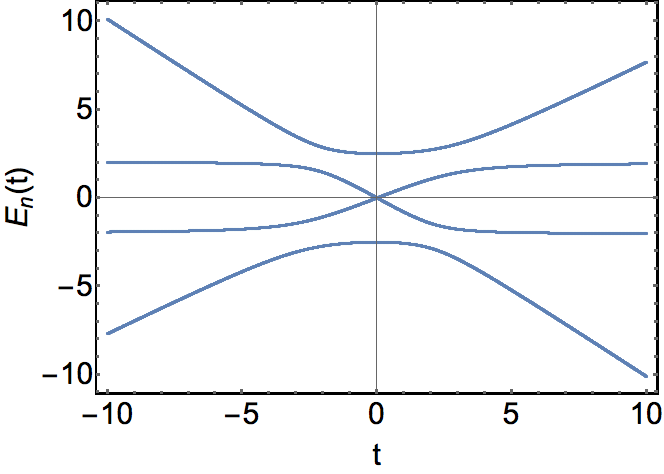}}
\hspace{-2mm}\vspace{-4mm}
\caption{Eigenvalues of the Hamiltonian (\ref{bham1}) as functions of $t$. Parameters: $\varepsilon=2$, $\beta_1=1$, $\beta_4=0.75$, $\gamma=0.7$, $g=0.8$.  There is an exact crossing point at time $t=0$ at the place of crossing of diabatic levels 1 and 4. This is a signature of Landau-Zener integrability.}
\label{bow-tie2}
\end{figure}

Our integrable model is shown in Fig.~\ref{bow-tie}. It is the 4-state bow-tie model, which is exactly solvable as described in Sec~\ref{prelim-anzatz}. Its Hamiltonian is
\be
\hat{H} =\left(
\begin{array}{cccc}
\beta_1 t & \gamma &  \gamma& 0 \\
\gamma & \varepsilon & 0& g \\
\gamma &0 & -\varepsilon &g  \\
0 &g &g & \beta_4 t
\end{array} \right),
\label{bham1}
\ee
where we assume $\varepsilon>0,\, \beta_1<0,\, \beta_4>0$. In Fig.~\ref{bow-tie2}, we show the dependence of eigenvalues of the Hamiltonian (\ref{bham1}) on time, treating $t$ as a parameter of the matrix. Note that at $t=0$ there is an exact crossing point that coincides with the crossing point of diabatic levels 1 and 4, which are not directly coupled to each other. According to \cite{six-LZ}, the presence of such crossing points is the signature of Landau-Zener integrability.


In order to derive transition probabilities, we first note that the Schr\"odinger equation of this model is invariant under the simultaneous application of the following commuting transformations:

(a) $t\rightarrow -t$;

(b) $a_2 \rightarrow a_3$, $a_3 \rightarrow a_2$;

(c) $a_1 \rightarrow -a_1$, $a_4 \rightarrow -a_4$.

So, if we take the conjugate transpose of the scattering matrix, exchange indexes 2 and 3 of its elements, and add minus signs at all elements with only one of the indexes being 1 or 4, the result will coincide with the original scattering matrix:
\begin{eqnarray}
\label{sc-bt1}
\nonumber\left(
\begin{array}{cccc}
S_{11}       &  S_{12}     &S_{13}   & S_{14}  \\
S_{21}  &S_{22}       & S_{23}         &   S_{24}  \\
S_{31}  & S_{32}        & S_{33} & S_{34} \\
S_{41}   & S_{42}  & S_{43} & S_{44}
\end{array}
\right)=\\
=\left(
\begin{array}{cccc}
S_{11}^*       &  -S_{31}^*     &-S_{21}^*   & S_{41}^*  \\
-S_{13}^*  &S_{33}^*       & S_{23}^*         &  - S_{43}^*  \\
-S_{12}^*  & S_{32} ^*       & S_{22}^* & -S_{42}^* \\
S_{14}^*   & -S_{34}^*  & -S_{24}^* & S_{44}^*
\end{array}
\right).
\end{eqnarray}
Comparing these two matrices, we can infer various relations; e.g., $S_{23}=S_{23}^*$ etc..  We then use such constraints to reduce the number of independent parameters of the scattering matrix
\begin{equation}
\hat{S}=\left(
\begin{array}{cccc}
X      &  S_{12}     &S_{13}   & S_{14}  \\
-S_{13}^*  &S_{22}       & A         &   S_{42}^*  \\
-S_{12}^*  & B               & S_{22}^* & S_{34} \\
S_{14}^*   & -S_{34}^*  & -S_{24}^* & Y
\end{array}
\right),
\label{sc-bt1}
\end{equation}
where
\be
A\equiv S_{23}, \quad B\equiv S_{32}
\label{ab1}
\ee
are introduced to emphasize that the corresponding matrix elements are real numbers. We also introduced the real-valued parameters
\be
X=e^{-\frac{2\pi |\gamma|^2}{|\beta_1|}}, \quad Y=e^{-\frac{2\pi |g|^2}{\beta_4}}, \quad Z\equiv\sqrt{XY}, 
\label{par4}
\ee
and we included HCs (\ref{be1}) and (\ref{be2}) in (\ref{sc-bt1})  explicitly.

The 2nd order HCs, (\ref{hhhc2}) and (\ref{hhhc21}), now lead to
\be
XS_{22}^*+S_{13}S_{12}^*=Z, \quad YS_{22}^*+S_{12}S_{12}^*=Z.
\label{hc-bt2}
\ee
In addition, two parallel levels in the model (\ref{bham1}) have the next-to-extremal slope, so we can apply the relations (\ref{ng11}) to them. Writing these conditions in terms of the elements of matrix (\ref{sc-bt1}) we find
\be
XB+|S_{12}|^2=0, \quad Y A+|S_{24}|^2=0.
\label{hc-bt3}
\ee

Consider now the unitarity constraint:
\be
[\hat{S}^{\dg} \hat{S} ]_{23}=0,
\label{un-bt1}
\ee
which explicitly reads
\be
S_{13}^*S_{12}+S_{22}(B+A) +S_{24}S_{24}^*=0.
\label{sss1}
\ee
With constraints (\ref{hc-bt2}), this simplifies to
\be
S_{22} \left( B+A-X-Y \right) +2Z=0.
\label{un-bt2}
\ee
Since the term $2Z$ and the factor $B+A-X-Y$ are always real, the condition (\ref{un-bt2}) can be generally satisfied only if $S_{22}$ is also real. This fact has profound consequences because the expressions (\ref{hc-bt2}) can now be converted to constraints that do not involve phases of scattering matrix elements. Indeed, rearranging terms and taking absolute value squared on both sides of these equations we find
\be
P_{12} P_{13} = (Z-XS_{22})^2,\quad P_{24}P_{24}=(Z-YS_{22})^2.
\label{hc-pr1}
\ee
Two more useful constraints follow from $[\hat{S}^{\dg} \hat{S}]_{22}=[\hat{S}^{\dg} \hat{S} ]_{33}=1$, which can be written in components as
\begin{eqnarray}
\label{hc-pr2}
P_{12}+B^2+S_{22}^2+P_{34}&=&1,\\
\label{hc-pr3}
P_{13}+S_{22}^2+A^2+P_{24}&=&1.
\end{eqnarray}
We can solve  Eqs.~(\ref{hc-pr1})-(\ref{hc-pr3}) for $P_{12}$, $P_{13}$, $P_{34}$, $P_{24}$ in terms of real parameters $A$, $B$, $S_{22}$. Substituting them back into (\ref{hc-bt3}) we obtain two constraints:
\begin{eqnarray}
\label{hc-pr5}
-XB&=& \frac{(Z-XS_{22})^2}{1-A^2-S_{22}^2+YA}, \\
\label{hc-pr6}
-YA&=& \frac{(Z-YS_{22})^2}{1-B^2-S_{22}^2+XB}.
\end{eqnarray}
Together with Eq.~(\ref{un-bt2}), Eqs.~(\ref{hc-pr5})-(\ref{hc-pr6}) provide three equations for three unknown variables $A$, $B$, and $S_{22}$. Since the equations are nonlinear, they generally have more than one solution.

From (\ref{hc-bt3}) follows that $A$ and $B$ are negative. Hence, from (\ref{un-bt2}) it follows that $S_{22}$ is positive. By restricting our search to only physical ranges $A,B \in (-1,0)$ and $S_{22}\in (0,1)$, we find
\be
A=Y-1, \quad B=X-1,\quad S_{22}=Z.
\label{hc-sol1}
\ee
Substituting this into (\ref{hc-pr1})-(\ref{hc-pr3}) and using other constraints of the type (\ref{dstoch}), we restore the exact solution of the 4-state bow-tie model.
\begin{widetext}
\begin{eqnarray}
\label{pbt}
\hat{P} =
\left(
\begin{array}{cccc}
X^2& X(1-X) &  Y(1-X)& (1-X)(1-Y) \\
(1-X)Y & XY & (1-Y)^2& Y(1-Y) \\
X(1-X) &(1-X)^2 & XY &X(1-Y)  \\
(1-X)(1-Y) &X(1-Y) &Y(1-Y) & Y^2
\end{array} \right),
\end{eqnarray}
\end{widetext}
which coincides with predictions of the semiclassical ansatz from Sec.~\ref{prelim-anzatz}.

\subsection{Pseudo bow-tie model}
\label{4pbtm}
\begin{figure}
\scalebox{0.1}[0.1]{\includegraphics{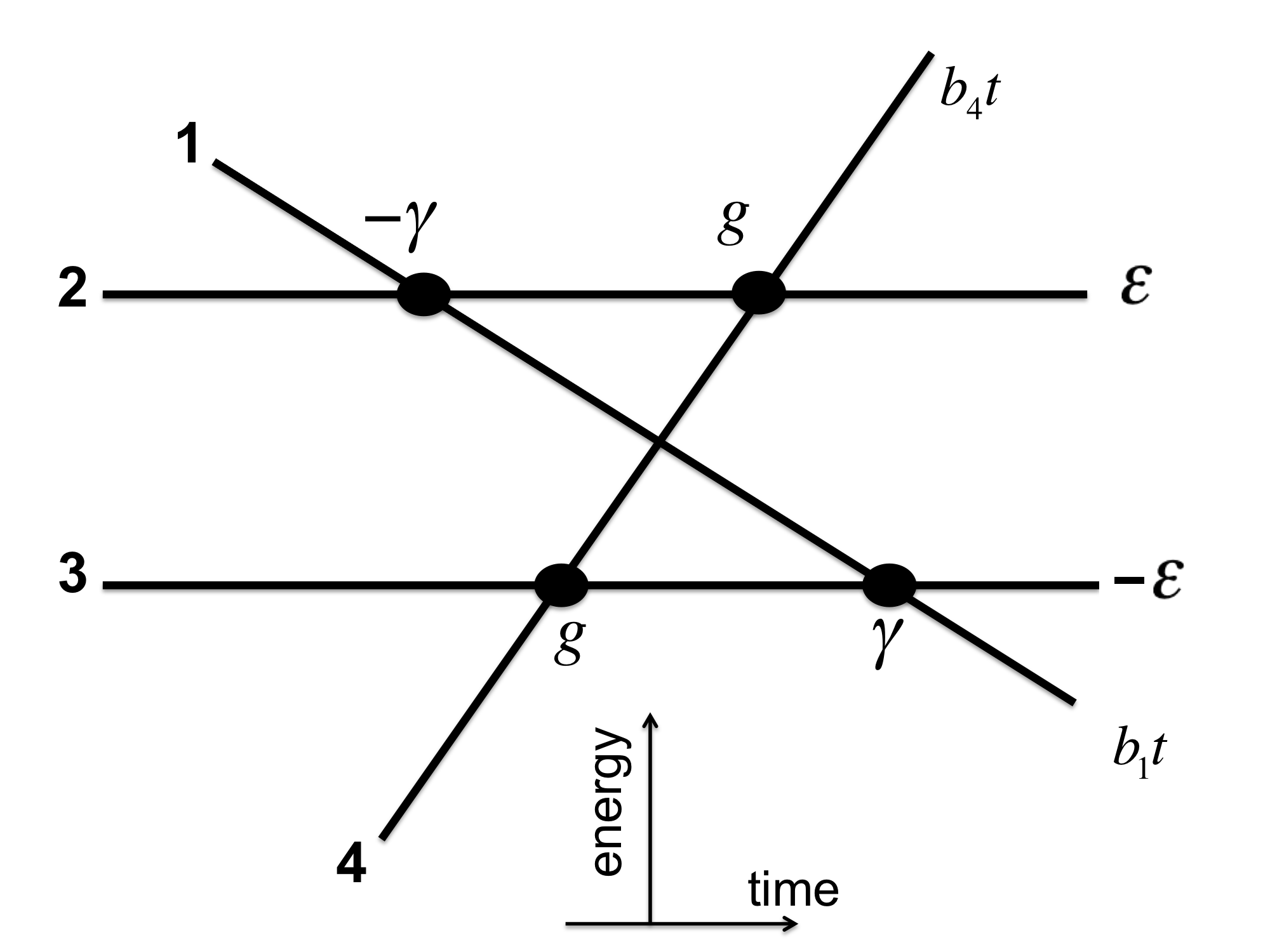}}
\hspace{-2mm}\vspace{-4mm}
\caption{ Diabatic levels and parameters of the nonintegrable model with the Hamiltonian (\ref{ham7}).}
\label{four-fig}
\end{figure}
\begin{figure}
\scalebox{0.28}[0.28]{\includegraphics{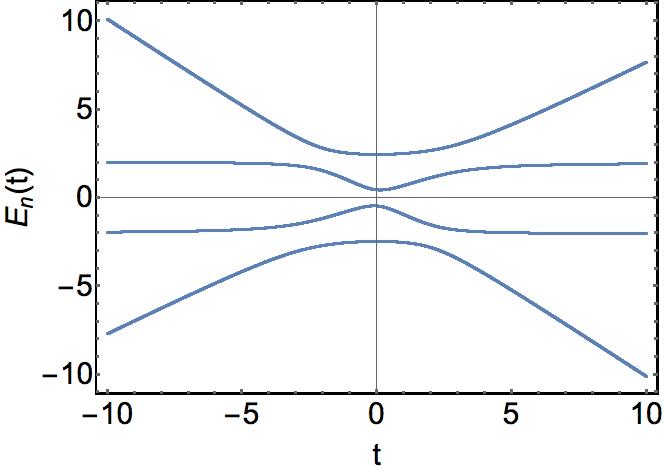}}
\hspace{-2mm}\vspace{-4mm}
\caption{Eigenvalues of the Hamiltonian (\ref{ham7}) as functions of $t$. Parameters are the same as in Fig.~\ref{bow-tie2}. Although according to Fig.~\ref{four-fig} diabatic levels 1 and 4 are not directly coupled to each other, there is a gap opening  near their crossing. This indicates that the model (\ref{ham7}) is not integrable.}
\label{four-fig22}
\end{figure}

Consider a four-state system with the Hamiltonian
\be
\hat{H} =\left(
\begin{array}{cccc}
\beta_1 t & -\gamma &  \gamma& 0 \\
-\gamma & \varepsilon & 0& g \\
\gamma &0 & - \varepsilon &g  \\
0 &g &g & \beta_4 t
\end{array} \right).
\label{ham7}
\ee
Its diabatic level diagram is shown in Fig.~\ref{four-fig}. We will assume the same choice of parameters as in model (\ref{bham1}). Model (\ref{ham7}) is different from model (\ref{bham1}) only by a sign change at one of the coupling constants. Figure~\ref{four-fig22} shows that, although diabatic levels 1 and 4 do not interact with each other directly, corresponding eigenvalues of the Hamiltonian (\ref{ham7}) experience avoided level crossings near the point $t=0$. Such a behavior is the signature of the breakdown of the model's integrability \cite{six-LZ}.

Nevertheless, model (\ref{ham7}) has almost the same discrete symmetry as the bow-tie model (\ref{bham1}). Namely, the Schr\"odinger equation with the Hamiltonian (\ref{ham7}) remains the same after  application of three mutually commuting operations:

(a) change of sign of time $t\rightarrow -t$;

(b) change of sign of one  amplitude: $a_4 \rightarrow - a_4$;

(c) renaming amplitudes of two parallel levels: $a_2 \rightarrow a_3$ and $a_3 \rightarrow a_2$.

Repeating the same steps as for the previously considered bow-tie model, we can parametrize the scattering matrix as
\be
\hat{S} = \left(
\begin{array}{cccc}
X& S_{12} & S_{21}^*          & S_{14} \\
S_{21} & S_{22} & A            & S_{24} \\
S_{12}^* &B      & S_{22}^* & S_{34} \\
-S_{14}^* & -S_{34}^* & -S_{24}^* & Y
\end{array} \right).
\label{sc4}
\ee
There are, in total, 8 unknown scattering amplitudes in (\ref{sc4}).  Considering the symmetries of (\ref{sc4}), only four constraints of the type (\ref{dstoch}) are expected to be independent. So for full integrability, we require four extra constraints on the transition probabilities. Second order HCs written in terms of the parametrization in (\ref{sc4}) read
\be
XS_{22}-S_{21}S_{12}=Z, \quad YS_{22}^* +S_{24}^*S_{34}=Z.
\label{cons41}
\ee
The analog of Eq.~(\ref{hc-bt3}) here is
\be
XB-|S_{12}|^2=0, \quad YA+|S_{24}|^2=0.
\label{ng2}
\ee

Considering equation $[\hat{S} \hat{S}^{\dg} ]_{23}=0$, we find
\be
S_{12}S_{21}+S_{22}(B+A) +S_{34}^*S_{24}=0,
\label{sss2}
\ee
which is the analog of Eq.~(\ref{sss1}) for the bow-tie model. So far, Eqs.~(\ref{cons41})-(\ref{sss2}) look very similar to their counterparts in the bow-tie model. Problems start when we substitute Eqs.~(\ref{cons41}) into (\ref{sss2}). We find then
\be
S_{22} \left[ X-Y+B+A \right]=0.
\label{sss3}
\ee
In comparison to the analogous Eq.~(\ref{un-bt2}), now the free term that depended on $Z$  has canceled out. As a result, we cannot conclude that $S_{22}$ is purely real and hence cannot plug it back to (\ref{cons41}) in order to derive additional constraints on transition probabilities. In fact, Eq.~({\ref{sss3}) tells nothing about $S_{22}$ now.

In principle, Eq.~(\ref{sss3}) still means a useful constraint on real parameters:
\be
B+A+X-Y=0.
\label{ba-cons}
\ee
Moreover, there is still one constraint on probabilities that can be derived from (\ref{cons41}). For this, we move terms with $S_{22}$ to the right hand side and take the absolute value squared on both sides.
Eliminating  ${\rm Re} (S_{22})$ we find a nonlinear relation for probabilities
\be
XY(X-Y)(P_{22}-1)=YP_{21}P_{12}-XP_{24}P_{34}.
\label{cons42}
\ee
One can still hope that Eqs.~(\ref{ng2}),~(\ref{ba-cons}), and (\ref{cons42}) provide the four missing constraints; however, our studies showed that this is not the case. They are not independent and only reduce the number of unknown parameters in the transition probability matrix to two.
However, when used altogether with four independent constraints (\ref{dstoch}), they lead to some  useful relations.  For example, if we set $P_{12}$ and $P_{22}$ as independent parameters then we find
\begin{eqnarray}
\label{con43}
\nonumber P_{32}&=&(P_{12}/X)^2,\\
\nonumber P_{23}&=&(Y-X-P_{12}/X)^2,\\
  \nonumber P_{24}&=&P_{43}=\frac{(P_{12}+X(X-Y))Y}{X}, \\
P_{14}&=&P_{41}=P_{22}+\frac{(P_{12}+X^2)(P_{12}-XY)}{X^2}.
\end{eqnarray}

\begin{figure}
\scalebox{0.28}[0.28]{\includegraphics{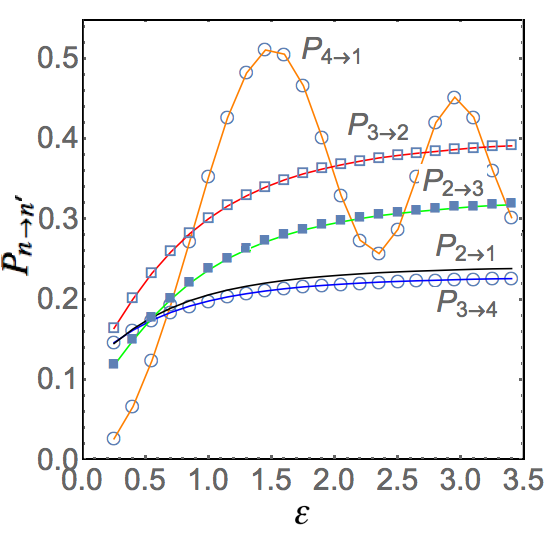}}
\hspace{-2mm}\vspace{-4mm}
\caption{(Color online) Numerical test of Eqs.~(\ref{con43}). Parameter $\varepsilon$  is the half-distance between parallel levels 2 and 3. Discrete points correspond to transition probabilities that were obtained directly from numerical simulations of quantum mechanical evolution  with the Hamiltonian (\ref{ham7}).  Solid curves
are predictions of formulas in (\ref{con43}) that take numerically obtained probabilities $P_{12} \equiv P_{2\rightarrow 1}$ and $P_{22} \equiv P_{2\rightarrow 2}$ as an input. Parameters for simulations: $\gamma=0.37$, $g=0.45$, $\beta_1=-1.$, $\beta_2=1.25$. Evolution proceeds from $t=-2000$ to $t=2000$ with a time step $dt=0.00005$. Simulation algorithm is described in Ref.~\cite{cQED-LZ}.}
\label{four-fig2}
\end{figure}
In Fig.~\ref{four-fig2}, we tested relations (\ref{con43}) by simulating evolution with the Hamiltonian (\ref{ham7}) numerically from large negative to large positive times. We found perfect agreement with our theory. Figure~\ref{four-fig2} demonstrates clearly that transition probabilities in the nonintegrable model are quite sensitive to the distance between parallel levels. In contrast, the solution of the integrable bow-tie model in Eq.~(\ref{pbt}) does not depend on $\varepsilon$ at all.

\section{Discussion}

Multistate Landau-Zener model (MLZM) has provided unusually many nonperturbative exact results. Some of them correspond to interactions among only a few states \cite{multiparticle,six-LZ, four-LZ},  others describe truly many-body mesoscopic dynamics \cite{cQED-LZ}.  In terms of complexity, DO and bow-tie models \cite{do,bow-tie} stay somewhere in between. It has always been puzzling why all such seemingly different systems produce  simple final results with many common properties. The discovery of the Brundobler-Elser formula in \cite{be} provided an important piece to this puzzle and generated a lot of interest.

Here, we showed that the Brundobler-Elser formula is actually only one of many nontrivial exact ``hierarchy constraints" (HC) on the scattering matrix of an arbitrary model of the type (\ref{mlz}). The effect of higher than 1st order HCs on transition probabilities is, however, not straightforward to understand in detail. We showed that HCs can lead to novel nonlinear relations among transition probabilities,
such as Eq.~(\ref{ch22}) for an arbitrary chain model and Eq.~(\ref{ng21}) for models with a band of parallel diabatic levels. 
 In combination with other symmetries, these relations can lead to considerable simplifications. Thus, using new HCs, we were able to prove expressions for transition probabilities of the spin-3/2 model that were conjectured in \cite{four-LZ}.

It is also encouraging that all known simple solvable systems, such as the three-state chain, DO, and the four-state bow-tie models can be fully solved using elementary symmetries and HCs. We argued that even such a general result as the no-go rule in MLZM \cite{no-go} is one of the consequences of HCs. Moreover, HCs depend only on specific parameter combinations that appear in all known exactly solved MLZMs with a finite number of levels.
All such observations point to the possibility that HCs are responsible for the phenomenon of integrability in MLZM. 

We would like to conclude with raising questions about MLZM that still need resolution. First, even in combination with other symmetries, HCs lead generally to a set of  strongly nonlinear algebraic equations. We showed how to resolve such equations in relatively simple situations but how to use even more nonlinear constraints remains an open problem. Second,  both HCs and solutions obtained by applying  the  semiclasscial ansatz, which we described in Sec.~\ref{prelim-anzatz}, depend on the same combinations of parameters.  This indicates that HCs can be responsible for the validity of the semiclassical ansatz. It remains unclear whether this is true and whether some other form of an exact solution is possible. 
Third, it is not clear what type of additional symmetry can be responsible  for most complex solvable cases such as in Ref.~\cite{cQED-LZ}. For example, quantum groups are under suspect \cite{chenLZ}. 

One unexplained signature of the Landau-Zener integrability is the presence of specific exact crossings of eigenvalues of the Hamiltonian \cite{com-partner,armen}, such as the one shown in Fig.~\ref{bow-tie2} for the four-state bow-tie model. We showed in Sec.~\ref{4btm} that, almost miraculously, all independent constraints play in synergy in this case, fixing the values of all transition probabilities. On the other hand, we showed in Sec.~\ref{4pbtm} that in a very similar model without such an eigenvalue crossing, phases of the scattering matrix cannot be eliminated from important equations. It is impossible then to close these equations for probability variables only.


In Ref.~\cite{sinitsyn-chemLZ}, the role of one type of  exact crossing points of adiabatic energy levels was completely understood within the context of multistate Landau-Zener problem. It was found that such points lead to specific constraints on the scattering matrix even for models that are not completely integrable. Thus, it seems that  exact crossing points play some important role, which is the next  key puzzle that should be resolved in order to connect numerous observations about MLZM into a coherent theory of integrable explicitly time-dependent quantum systems.

Apart from making an insight into the origin of the Landau-Zener integrability, HCs set an unusual example of an exact result that describes evolution with an arbitrary Hamiltonian of some very broad type. 
 Indeed, in $\hat{H}(t)=\hat{A}+\hat{B}t$, matrices $\hat{A}$ and $\hat{B}$ are  practically arbitrary by definition. Thus, at $t=0$, the Hamiltonian $\hat{H}=\hat{A}$ can describe a macroscopic system with strongly nonlinear and chaotic behavior. However, applying linearly growing fields and then measuring scattering amplitudes between asymptotically simple microstates it becomes possible to extract some simple combinations of  the matrix $\hat{A}$ elements. Having such properties, HCs can  extend  previously discussed applications of the Brundobler-Elser formula in physics of decoherence \cite{coher} and dynamic passage through a quantum phase transition \cite{chain}.
 
 The  example of HCs in MLZM raises questions about existence of similar results beyond the linearly time-dependent Hamiltonian. Such extensions are certainly possible within the bigger class of so-called LZC-systems that include terms  decaying as $\sim 1/t$ with time. Analogs of the Brundobler-Elser and the no-go relations  have been already derived for such models in \cite{sinitsyn-14pra}, so corresponding extensions of HCs can be found along the same path as in  the present article. The question ``what is the most general  evolution equation leading to HCs" is important  but still open. We hope that our article will stimulate interest  also  inside the mathematical community to resolve it.

\appendix

\section{Scattering Matrices of Landau-Zener Models}
\label{sec:S-LZ-linear-algebra}

In this appendix we provide a definition of the scattering matrix $S_{nn'}$, associated with a multistate Landau-Zener (LZ) problem. This is done mostly for the sake of introducing a convention, which is needed since the off-diagonal elements of $\hat{S}$ are defined up to phase factors that do not depend on the model parameters, and a convention is needed to fix the aforementioned ambiguity.

As stated in section~\ref{sec:intro}, a multistate LZ-problem is a system of linear equations, given by Eq.~(\ref{mlz}).
Diagonalizing the matrix $\hat{B}$, using a basis set $(e_1, \ldots, e_N)$ we obtain a set of real eigenvalues $\beta_1, \ldots, \beta_N$, referred to as the diabatic slopes. In case of degeneracy there is ambiguity in the choice of the basis set. However in this case we can consider the vector subspace of the eigenmodes that correspond to a degenerate eigenvalue $\beta_k$, project the matrix $\hat{A}$ to this subspace, obtaining a  hermitian matrix, and further diagonalize the latter using a basis set. We assume that the eigenvalues of the aforementioned projection, referred to as diabatic energies, are non-degenerate, treating the latter condition as a restriction, needed to define a proper LZ-problem, i.e., a one that possesses a well-defined scattering matrix. Applying the above procedure to all degenerate levels of $\hat{B}$ we obtain an ordered basis set by applying the lexicographic ordering, introduced in section~\ref{sec:intro}, i.e., $i > j$, if $\beta_i > \beta_j$, or if $\beta_i = \beta_j$ then $\varepsilon_i < \varepsilon_j$. The obtained ordered basis set $(e_1, \ldots, e_N)$ is referred to as the diabatic basis set; it is defined with a minimal ambiguity of choosing phases of the normalized  elements of the basis.

We further introduce the adiabatic phases $\varphi_k(t)$, associated with the diabatic states $e_k$
\begin{eqnarray}
\label{adiabatic-phase} \varphi_{k}(t) &=& -\frac{\beta_{k}}{2}t^{2} - \varepsilon_{k}t - \frac{\eta_k}{2}\ln(t^{2} + 1), \nonumber \\ \eta_k &=& \sum_{l}^{l \ne k}\frac{|g_{kl}|^{2}}{\beta_k - \beta_l}.
\end{eqnarray}
The reason for such a choice of the phases is the following. At $t \rightarrow \pm \infty$ the system evolves adiabatically. Since the difference between the adiabatic and diabatic states tends to zero as $\sim |t|^{-1}$ at $t \rightarrow \pm \infty$, we can represent $N$ linearly independent solutions $\Psi_{k}^{\pm}(t)$ asymptotically at $t \rightarrow \pm \infty$, respectively,
\begin{eqnarray}
\label{adiabatic-phase-2} \Psi_{k}^{\pm}(t) \sim e_{k}e^{i\varphi_{k}^{\pm}(t)}, \;\;\;  \varphi_{k}^{\pm}(t) = -\int_{t_{\pm}}^{t}d\tau \bar{\varepsilon}_{k}(\tau),
\end{eqnarray}
with $t_{\pm}$ sitting in the corresponding adiabatic regions, and $\bar{\varepsilon}_{k}(\tau)$ being the adiabatic energies. Applying the standard perturbation theory
\begin{eqnarray}
\label{adiabatic-energy}\bar{\varepsilon}_{k}(t) \sim \beta_{k}t + \varepsilon_{k} + \frac{\eta_{k}}{t},
\end{eqnarray}
where we neglected the higher-order terms, since they tend to zero after time integration in the adiabatic regions. Substituting Eq.~(\ref{adiabatic-energy}) into Eq.~(\ref{adiabatic-phase-2}) we obtain the asymptotic expressions for the adiabatic phase in the adiabatic regions
\begin{eqnarray}
\label{adiabatic-phase-3} \varphi_{k}^{\pm}(t) = -\frac{\beta_{k}}{2}t^{2} - \varepsilon_{k}t - \eta_k\ln|t| + c_{k}^{\pm},
\end{eqnarray}
and setting the integration constants $c_{k}^{\pm} = 0$ to zero, which actually determines our convention, we arrive at the expression in Eq.~(\ref{adiabatic-phase}) that has the same asymptotic form as in Eq.~(\ref{adiabatic-phase-3}), while being defined on the whole real axis.

Since $\Psi_{k}^{\pm}(t)$ form two basis sets in the $N$-dimensional vector space of the solutions of the LZ-problem [Eq.~(\ref{mlz})], the scattering matrix can be defined as the transformation from one to another:
\begin{eqnarray}
\label{define-S} \Psi_{n}^{+}(t) = \sum_{n'}S_{nn'}\Psi_{n'}^{-}(t),
\end{eqnarray}
and note that by definition $\hat{S}$ does not depend on time, whereas $\Psi_{k}^{\pm}(t)$ are well defined for all real $t$, and also can be uniquely analytically continued to all complex values of $z$, while they have a simple asymptotic form at $t \rightarrow \pm \infty$. Since the solutions of Eq.~(\ref{mlz}) can be explicitly expressed in terms of the evolution operator
\begin{eqnarray}
\label{define-U} \hat{U}(t, t') = T\exp\left(-i\int_{t'}^{t}d\tau \hat{H}(\tau)\right),
\end{eqnarray}
we arrive at an explicit expression for the scattering matrix
\begin{eqnarray}
\label{S-explicit} S_{nn'} = \lim_{t \rightarrow \infty}\lim_{t' \rightarrow -\infty}U_{nn'}(t, t')e^{-i(\varphi_{n}(t) - \varphi_{n'}(t'))}.
\end{eqnarray}

The expression for the scattering matrix in terms of the evolution counterpart allows the scattering matrices of the composite LZ-problems to be derived based on the corresponding properties of the evolution matrices, the latter being dealt using standard linear algebra.

While the adiabatic phases $\varphi_k(t)$ do not enter the expressions for the scattering matrices of the composite (tensor-product and exterior-product) LZ-problems, considered in appendix~\ref{sec:LZ-linear-algebra} [Eqs.~(\ref{S-composite}) and (\ref{S-n-fermion})], and need to be carefully considered just for justification purposes, they play an important and explicit role in certain symmetry properties of the LZ-problems. In particular, linear transformations of time $t = \lambda t' + t_{0}$ that change both the time scale and the position of its reference zero point. By switching to the new variables in Eq.~(\ref{mlz}) we obtain the transformed value of the LZ-problem parameters
\begin{eqnarray}
\label{transform-parameters} \beta_{k}' = \lambda^{2}\beta_{k}, \;\; \varepsilon_{k}' = \lambda(\varepsilon_{k} + \beta_{k}t_{0}), \;\; g_{kr}' = \lambda g_{kr}.
\end{eqnarray}
A straightforward computation yields for $t \rightarrow \pm \infty$:
\begin{eqnarray}
\label{phase-difference} \varphi_{k}(t) - \varphi_{k}'(t') \sim \frac{\beta_{k}t_{0}^{2}}{2} + \varepsilon_{k}t_{0} + \eta_{k}\ln|\lambda| \equiv \zeta_{k},
\end{eqnarray}
which yields the following relation between $\hat{S} = \hat{S}(\beta, \varepsilon, g)$ and $\hat{S}' = \hat{S}(\beta', \varepsilon', g')$:
\begin{eqnarray}
\label{S-under-reparamet-t} S_{nn'}' = S_{nn'}e^{i(\zeta_{n} - \zeta_{n}')}.
\end{eqnarray}

\section{Linear Algebra and Composite Models}
\label{sec:LZ-linear-algebra}

In this appendix we will interpret the procedure of building composite LZ-problems, including the models of non-interacting fermions, applied in section~\ref{derivation} to derive the constraints, as linear algebra in the space of multistate LZ-problems.

Consider two LZ-problems in vector spaces $V_1$ and $V_2$ of dimensions $N_1$ and $N_2$, with the Hamiltonians $\hat{H}_1(t) = \hat{A}_1 + \hat{B}_{1}t$ and $\hat{H}_2(t) = \hat{A}_2 + \hat{B}_{2}t$ respectively. As outlined above the vector spaces are equipped with the preferred ordered diabatic basis sets $(e_{1}^{(1)}, \ldots e_{N_1}^{(1)})$ and $(e_{1}^{(2)}, \ldots e_{N_2}^{(2)})$. The composite LZ-problem, also referred to as the tensor product of the above two, has the space $V = V_{1} \otimes V_{2}$, of dimension $N = N_{1}N_{2}$ equipped with the preferred basis set, represented by $e_{ks} = e_{k}^{(1)} \otimes e_{s}^{(2)}$, with the Hamiltonian
\begin{eqnarray}
\label{H-composite} \hat{H}(t) &=& \hat{H}_1(t) \otimes \hat{I} + \hat{I} \otimes \hat{H}_2(t), \nonumber \\ \hat{A} &=& \hat{A}_1 \otimes \hat{I} + \hat{I} \otimes \hat{A}_2, \nonumber \\ \hat{B} &=& \hat{B}_1 \otimes \hat{I} + \hat{I} \otimes \hat{B}_2.
\end{eqnarray}
It follows from Eq.~(\ref{mlz}) that $\hat{U}(t, t') = \hat{U}^{(1)}(t, t') \otimes \hat{U}^{(2)}(t, t')$, which, combined with the obvious property $\varphi_{kr}(t) = \varphi_{k}^{(1)}(t) + \varphi_{r}^{(2)}(t)$ immediately extends the product property of the evolution operator to the scattering matrices
\begin{eqnarray}
\label{S-composite} \hat{S} = \hat{S}^{1} \otimes \hat{S}^{(2)}, \;\; S_{kr, k'r'} = S_{kk'}^{(1)}S_{rr'}^{(2)}.
\end{eqnarray}

If we consider a tensor product of a space with itself, and further an iterated $n$-fold tensor product, we can restrict ourselves to its completely antisymmetric components called the $n$-th exterior degree, denoted $\bigwedge^n V$, which is equipped with a preferred basis set, given by $e_{k_1 \ldots k_n} = e_{k_1} \wedge \ldots \wedge e_{k_n}$ in terms of the preferred basis set of $V$, associated with a LZ-problem, with the wedge denoting an antisymmetric product. In language of physics, $\bigwedge^n V$ represents $n$-fermion states of the system. For a linear map $f: V \rightarrow V$ we consider its $n$-th exterior power $\wedge^n f: \bigwedge^n V \rightarrow \bigwedge^n V$ defined by
\begin{eqnarray}
\label{define-ext-power-map} \wedge^n f(u_1 \wedge \ldots \wedge u_n) = f(u_1) \wedge \ldots \wedge f(u_n).
\end{eqnarray}
Then the result obtained in section~\ref{derivation}, reformulated in linear algebra terms means $\hat{U}^{(n)}(t, t') = \wedge^n \hat{U}(t, t')$, with $\hat{U}^{(n)}(t, t')$ being the evolution operator of the $n$-fermion LZ-problem. Combining it with the property
\begin{eqnarray}
\label{adiabatic-phase-composite} \varphi_{k_1 \ldots k_n}(t) = \varphi_{k_1}(t) + \ldots + \varphi_{k_n}(t),
\end{eqnarray}
inherited from the tensor product property of the adiabatic phases, considered above in the context of the tensor products of LZ-problems, we arrive to the following property of the $n$-fermion scattering matrix
\begin{eqnarray}
\label{S-n-fermion} \hat{S}^{(n)} = \wedge^{n}\hat{S},
\end{eqnarray}
and in particular this means that the matrix elements of $S^{(n)}$ are given by the determinants of the  corresponding $n \times n$ minors of the $N \times N$ scattering matrix $\hat{S}$ of the original LZ-problem.

\section{Example of a Composite Model}
\label{reduce-ex}
\begin{figure}
\scalebox{0.14}[0.14]{\includegraphics{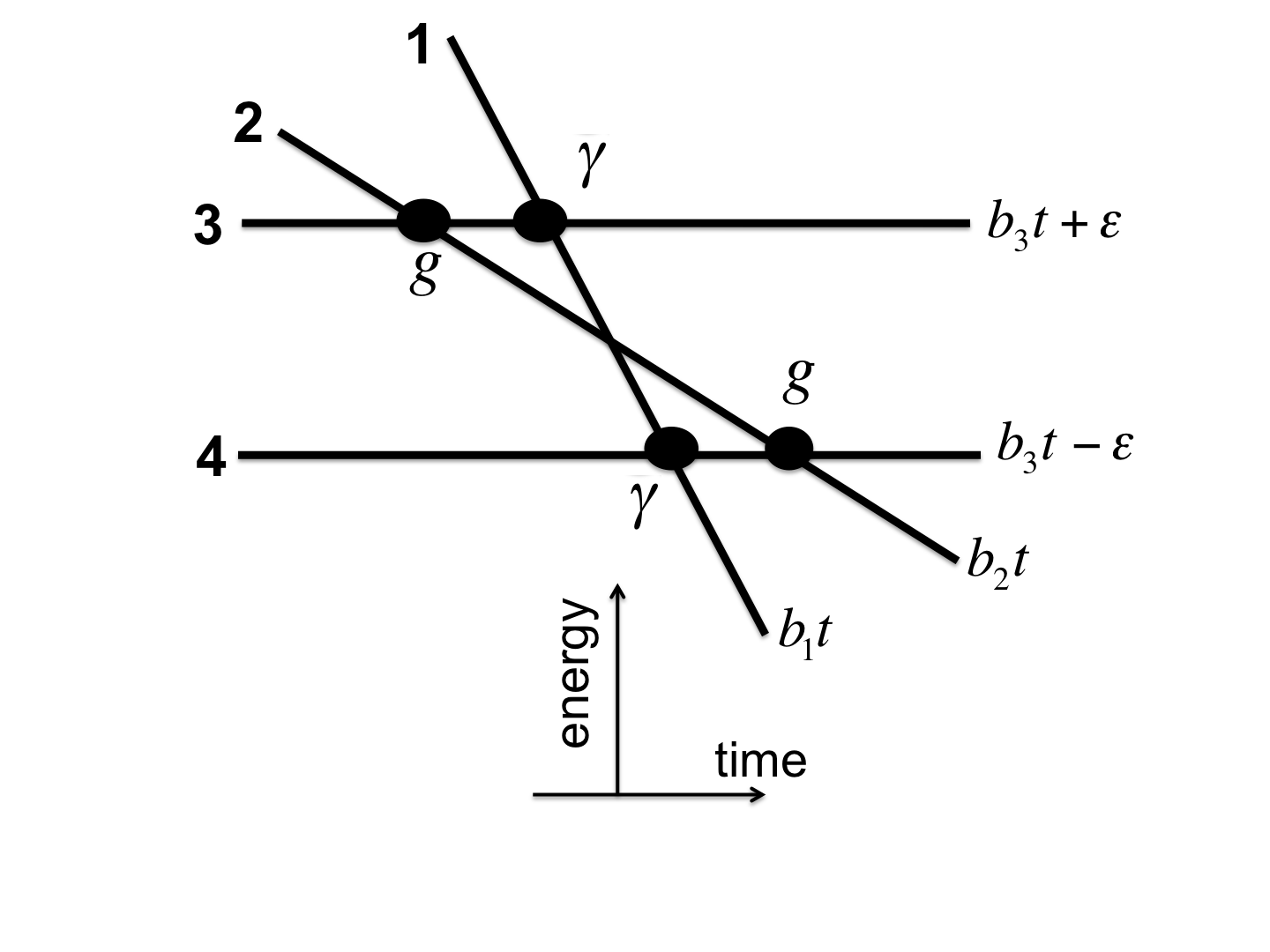}}
\hspace{-2mm}\vspace{-4mm}
\caption{ Diabatic level dagram of the 4-state Hamiltonian corresponding to the single particle sector of the model (\ref{ham2}) at $b_1>b_2>b_3$.}
\label{six33-fig}
\end{figure}
It was shown in \cite{multiparticle} that one can generate new solvable multistate Landau-Zener models from already solved ones.
To do this, one should assume that a known solvable model describes evolution of either fermionic or bosonic operators in the Heisenberg picture.
Switching to the Fock space, i.e. to evolution of the state amplitudes, we find then the matrix Hamiltonian that depends on the number of particles. When restricted to the single particle sector, this Hamiltonian describes the original model but multiparticle sectors look more complex.

The goal of this appendix is to support our discussion in Sec.~\ref{derivation} with a specific example of this procedure. In order not to overlap with Ref.~\cite{multiparticle}, we choose a model that was not studied there. As a byproduct, we will provide the proof of the six-state model solution that was conjectured in Ref.~\cite{six-LZ} based on integrability conditions.

Consider the secondary quantized Hamiltonian of four interacting fermions
\begin{eqnarray}
\label{ham2}
\hat{H}&=&\beta_1 t \hat{a}^{\dg} \hat{a} + \beta_2t \hat{b}^{\dg} \hat{b} + (\beta_3 t +e)\hat{c}_1^{\dg}\hat{c}_1 +(\beta_3 t -\\
\nonumber && -e)\hat{c}_2^{\dg} \hat{c}_2 + [g\hat{a}^{\dg} (\hat{c}_1+\hat{c}_2) +\gamma \hat{b}^{\dg} (\hat{c}_1+\hat{c}_2) + {\rm h.c.}],
\end{eqnarray}
with constant parameters $\beta_1$, $\beta_2$, $\beta_3$, $g$, $\gamma$, $e$. Heisenberg evolution equation for operators then reads:
\be
i\frac{d}{dt}
\left(
\begin{array}{c}
\hat{a}\\
\hat{b}\\
\hat{c}_1\\
\hat{c}_2
\end{array} \right)=\left(
\begin{array}{cccc}
\beta_1 t & 0 & g & g \\
0 & \beta_2 t & \gamma & \gamma \\
g &\gamma & \beta_3 t +e & 0 \\
g &\gamma & 0 & \beta_3 t -e
\end{array} \right) \left(
\begin{array}{c}
\hat{a}\\
\hat{b}\\
\hat{c}_1\\
\hat{c}_2
\end{array} \right).
\label{haiz2}
\ee
Equation~(\ref{haiz2}) has the same form as the Schr\"odinger equation for amplitudes in the 4-state bow-tie model, whose scattering matrix is known. Since Eq.~(\ref{haiz2}) is linear in operators, we can write its solution in terms of the scattering matrix elements, $S_{ij}$ of the bow-tie model. We can read elements $S_{ij}$ directly from the diagram in Fig.~\ref{six33-fig} using the semiclassical ansatz defined in Sec.~\ref{prelim-anzatz}.
Let us denote
$$
p_1=e^{-2\pi \frac{g^2}{|b_3-b_2|}}, \quad p_2=e^{-2\pi \frac{\gamma^2}{|b_3-b_1|}}, \quad q_{1,2}=1-p_{1,2}.
$$
The scattering matrix for a single particle sector with $b_1<b_2<b_3$ then reads:
\begin{eqnarray}
\label{dsds}
&&\hat{S}=\\
\nonumber &&\left(
\begin{array}{cccc}
p_2& -\sqrt{p_2q_2q_1} & i\sqrt{p_1p_2q_2} & i\sqrt{q_2} \\
-\sqrt{p_2q_2q_1} & p_1+q_1q_2 & ip_2\sqrt{p_1q_1} & i\sqrt{p_2q_1} \\
i\sqrt{q_2} &i\sqrt{p_2q_1} & \sqrt{p_1p_2} & 0 \\
i\sqrt{p_2q_2p_1} & ip_2\sqrt{p_1q_2} & -q_1-p_1q_2 & \sqrt{p_1p_2}
\end{array} \right).
\end{eqnarray}

Let us now consider the sector of the model (\ref{ham2}) with two fermions. In the basis
\begin{eqnarray}
\label{basis-62}
 |1\ra &\equiv& \hat{a}^{\dg} \hat{b}^{\dg} |0\ra, \quad  |2\ra \equiv \hat{c}_1^{\dg} \hat{c}_2^{\dg} |0\ra, \quad  |3\ra \equiv \hat{a}^{\dg} \hat{c}_1^{\dg} |0\ra,  \\
\nonumber |4\ra &\equiv& \hat{a}^{\dg} \hat{c}_2^{\dg} |0\ra, \quad |5\ra \equiv \hat{b}^{\dg} \hat{c}_1^{\dg} |0\ra, \quad |6\ra \equiv \hat{b}^{\dg} \hat{c}_2^{\dg} |0\ra,
\end{eqnarray}
the Hamiltonian (\ref{ham2}) is a 6$\times$6 matrix:
\begin{widetext}
\begin{equation}
\hat{H}'=\left(
\begin{array}{cccccc}
(\beta_1+\beta_2) t   & 0                   &\gamma         		 & \gamma               & -g    & -g		\\
0                &  2\beta_3 t	& -g          		       & g                     &-\gamma 	&\gamma	 \\
\gamma                & -g      		& (\beta_1+\beta_3) t +e    		 &0                         & 0	&0	 \\
\gamma	   	& g&0 					&(\beta_1+\beta_3) t -e  & 0		& 0		\\			
-g         & -\gamma                 & 0   				& 0      		     & (\beta_2+\beta_3) t +e & 0	\\
-g		 & \gamma		&0					&0				&0 	& (\beta_2+\beta_3) t-e
\end{array}
\right).
\label{six-ham22}
\end{equation}

Up to state renumbering, model (\ref{six-ham22}) contains the six-state model in Ref.~\cite{four-LZ} as a special case. The scattering amplitudes in model (\ref{six-ham22}) can be derived now using Eqs.~(\ref{pr1}) and (\ref{dsds}). For example,
$
S'_{11}  = S_{11}S_{22}-S_{12}S_{21}=p_1p_2.
$
Taking $P'_{ij}\equiv |S'_{ij}|^2$ of such amplitudes, we find the matrix of transition probabilities of the composite six-state model:
\begin{equation}
\hat{P}'=\left(
\begin{array}{cccccc}
p_1^2p_2^2&0&p_1q_1p_2^2&p_2q_1&p_1p_2q_2& q_2 \\
0&p_1^2p_2^2&q_2&p_1p_2q_2&p_2q_1& p_1q_1p_2^2\\
p_2q_1&p_1p_2q_2&p_1p_2&q_2^2&0&p_2q_2q_1\\
p_1q_1p_2^2&q_2&p_2^2q_1^2&p_1p_2&p_2q_2q_1&0\\
q_2&p_1q_1p_2^2&0&p_2q_2q_1&p_1p_2&p_2^2q_1^2\\
p_1p_2q_2&p_2q_1&p_2q_2q_1&0&q_2^2&p_1p_2
\end{array}
\right).
\label{six-ham2}
\end{equation}
\end{widetext}
The matrix (\ref{six-ham2}) coincides, up to redefinition of state indexes, with the matrix in Eq.~(32) in Ref.~\cite{four-LZ}. This fact proves the conjecture made in that reference about the integrability of its six-state model.

\section{Test of a more complex constraint for independence of Eqs.~(\ref{h111})-(\ref{h222})}
\label{test}

Let us try to generate a new constraint by considering  the MLZM Hamiltonian $\hat{H}$, which we populate with two non-interacting fermions to generate a new matrix Hamiltonian. Let us apply the second level HC (\ref{hhhc2}) to the two-fermion scattering matrix:
 \begin{eqnarray}
  \label{HC-ad1}
 &&{\rm Det} \left(
 \begin{array}{cc}
 S_{12,12}& S_{12,13} \\
 S_{13,12}& S_{13,13}
 \end{array}
 \right) = e^{-\sum \limits _{k  \ne 1,2}\frac{\pi |g_{k1}|^2}{|\beta_1 - \beta_k|}}\times\\
 \nonumber &\times&e^{-\sum \limits _{k \ne 1,3}\frac{\pi |g_{k1}|^2}{|\beta_1 - \beta_k|}-
 \sum \limits _{k  \ne 1,2,3}\frac{\pi |g_{k2}|^2}{|\beta_2 - \beta_k|}-\sum \limits _{k  \ne 1,2,3}\frac{\pi |g_{k3}|^2}{|\beta_3 - \beta_k|}},
 \end{eqnarray}
 where indexes $ij$ mark two levels of the original model that are populated by fermions. On the other hand, Eq.~(\ref{pr1}) gives
 \begin{eqnarray}
  \label{HC-ad2}
\nonumber S_{12,12} &=& S_{11}S_{22}-S_{12}S_{21}, \,\,\, S_{12,13}=S_{11}S_{23}-S_{21}S_{13},\\
\nonumber S_{13,12} &=& S_{11}S_{32}-S_{31}S_{12}, \,\,\,  S_{13,13}=S_{11}S_{33}-S_{13}S_{31}.
 \end{eqnarray}
Substituting this into (\ref{HC-ad1}) we find a constraint that relates some of the amplitudes and parameters of the original model. However, it is straightforward to check that
 \be
 {\rm Det} \left(
 \begin{array}{cc}
 S_{12,12}& S_{12,13} \\
 S_{13,12}& S_{13,13}
 \end{array}
 \right) = S_{11}  {\rm Det} \left(
 \begin{array}{ccc}
 S_{11}& S_{12} & S_{13} \\
 S_{21}& S_{22} & S_{23}\\
 S_{31} & S_{32} &S_{33}
 \end{array}
 \right),
 \label{HC-ad3}
 \ee
so, constraint (\ref{HC-ad1}) is merely the result of a product of HCs (\ref{be1}) and (\ref{hhhc3}). Hence, the constraint (\ref{HC-ad1}) is a pure consequence of the hierarchy (\ref{h111}).

Finally, we note that  hierarchies (\ref{h111}) and (\ref{h222}) are not independent of each other because populating  $M$ lowest slope levels of an $N$-state system with noninteracting electrons is physically equivalent to populating $N-M$ highest slope levels with noninteracting  holes. While electrons produce HC  (\ref{h111}), holes produce HC  (\ref{h222}), so the second hierarchy follows from the first one via the particle-hole duality of fermionic systems. Therefore, we have generally only $N-1$ independent HCs in an arbitrary $N$-level model.

\section{Reduction of solvable models}
\label{reduction}

Here we show that the model in Sec.~\ref{spin32-sec} can be solved in an alternate way which we will call the {\it reduction}.  
Namely, we show that solution  of the 4-state model in Sec.~\ref{spin32-sec} can be derived from solution of a more complex  but previously solved 6-state bow-tie model.

Let us populate four diabatic levels of the Hamiltonian (\ref{ham44}) with two noninteracting fermions. The corresponding fermionic Hamiltonian reads:
\begin{eqnarray}
\label{ham4}
\nonumber \hat{H}=(\beta_1 t+e) \hat{a}^{\dg} \hat{a} &+&(-\beta_1t+e) \hat{b}^{\dg} \hat{b} + (\beta_2 t -e)\hat{c}^{\dg}\hat{c} +(-\beta_2 t \\
-e)\hat{d}^{\dg} \hat{d}
+ g[\hat{a}^{\dg} \hat{c}&+&\hat{b}^{\dg} \hat{d} +{\rm h.c.}] +\gamma[ \hat{a}^{\dg} \hat{d}-\hat{b}^{\dg} \hat{c} + {\rm h.c.}].
\end{eqnarray}
We then choose the basis
\begin{eqnarray}
\label{basis-63}
\nonumber |1\ra &\equiv& \hat{a}^{\dg} \hat{b}^{\dg} |0\ra, \quad  |2\ra \equiv \hat{c}^{\dg} \hat{d}^{\dg} |0\ra,   \\
  |3\ra &\equiv& \hat{a}^{\dg} \hat{c}^{\dg} |0\ra, \quad   |4\ra \equiv \hat{a}^{\dg} \hat{d}^{\dg} |0\ra,  \\
\nonumber \quad |5\ra &\equiv& \hat{b}^{\dg} \hat{c}^{\dg} |0\ra, \quad |6\ra \equiv \hat{b}^{\dg} \hat{d}^{\dg} |0\ra.
\end{eqnarray}

\begin{figure}
\scalebox{0.1}[0.1]{\includegraphics{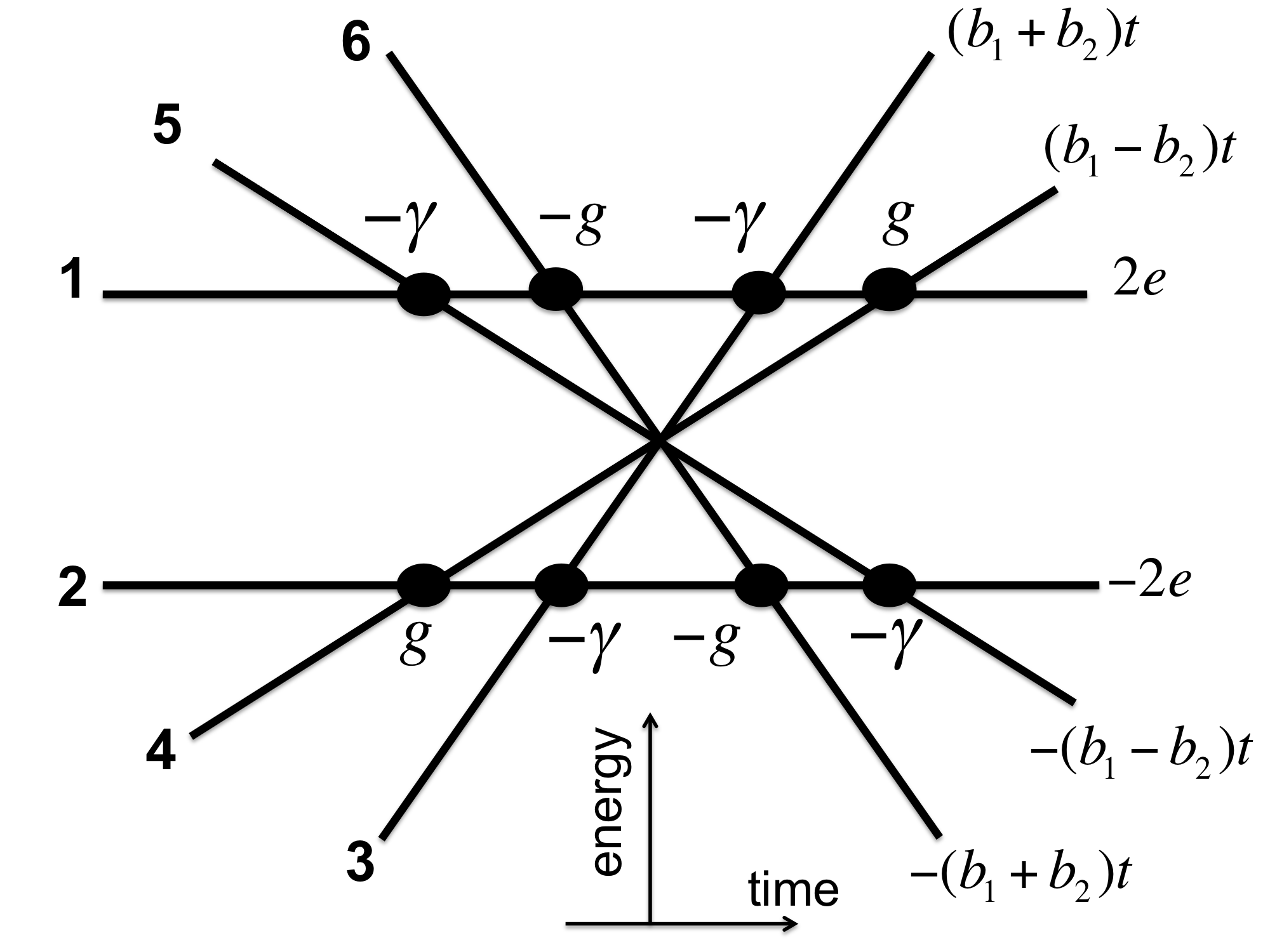}}
\hspace{-2mm}\vspace{-4mm}
\caption{ Diabatic levels of the Hamiltonian (\ref{ham4}) in the basis (\ref{basis-63}). }
\label{six3-fig}
\end{figure}
Instead of writing an explicit matrix, we illustrate the parameters of the corresponding 6$\times$6 Hamiltonian in Fig.~\ref{six3-fig}. It is clear that this model belongs to the class of bow-tie models shown in Fig~\ref{do-bt-fig}(b).

Let us derive the transition probability $P_{3\rightarrow 1}^{(4)} \equiv |S_{13}|^2$, where the upper index ``(4)"  marks the probabilities and indexes in the 4-state model (\ref{ham44}). The probability to find the fermion $\hat{a}$ after evolution with the Hamiltonian (\ref{ham4}) in the two-fermion sector, assuming that the system starts in state $|3\ra$, defined in (\ref{basis-63}), is given by
\be
\la 3| \hat{a}^{\dg}(+\infty) \hat{a}(+\infty) |3 \ra^{(6)} = P_{1\rightarrow 1}^{(4)}+P_{3 \rightarrow 1}^{(4)},
\label{prob1}
\ee
where the upper index ``(6)" means that we refer to the model in Fig.~\ref{six3-fig}. On the other hand, in terms of the Hamiltonian illustrated in Fig.~\ref{six3-fig}, such a probability is the sum of probabilities of transitions from level 3 in this figure to all levels that correspond to a filled fermion of type $\hat{a}$, which are levels 1, 3, and 4 in Fig.~\ref{six3-fig}. The bow-tie model solution is reproduced by a semiclassical ansatz described in Sec.~\ref{prelim-anzatz}. None of the transitions from level 3 to levels 1, 3, and 4 involve path interference, so we can readily read such transition probabilities:
\begin{eqnarray}
P_{3\rightarrow 3}^{(6)} &=& p_2^2, \\
P_{3\rightarrow 1}^{(6)} &=& p_2(1-p_2)p_1, \\
P_{3\rightarrow 4}^{(6)} &=& p_2(1-p_2)(1-p_1),
\label{prob2}
\end{eqnarray}
where $p_1$ and $p_2$ are defined in (\ref{p1p2}). We also know, from Eq.~(\ref{be1}), that $ P_{1\rightarrow 1}^{(4)}=p_1p_2$. We can now equate results of both ways to calculate  $\la 3| \hat{a}^{\dg}(+\infty) \hat{a}(+\infty) |3 \ra^{(6)}$:
\be
P_{3\rightarrow 1}^{(6)} +P_{3\rightarrow 3}^{(6)} +P_{3\rightarrow 4}^{(6)} =P_{1\rightarrow 1}^{(4)}+P_{3 \rightarrow 1}^{(4)},
\label{eq1}
\ee
which leads to
\be
P_{3 \rightarrow 1}^{(4)}=p_2(1-p_1),
\label{eq2}
\ee
which in turn coincides with the main result in Sec.~\ref{spin32-sec}. This approach to solve the spin-3/2 model demonstrates that some instances of already solved models can be themselves reducible to interesting and more compact systems like the spin-3/2 model.

{\it Acknowledgment}. This work was carried out under the auspices of the National Nuclear Security Administration of the U.S. Department of Energy at Los
Alamos National Laboratory under Contract No. DE-AC52-06NA25396.  V.Y.C. was supported by the National Science Foundation under Grant No. CHE-1111350. N.A.S. also thanks the support from the LDRD program at LANL.

\end{document}